\documentclass[12pt]{article}
\usepackage{amssymb}
\usepackage{axodraw}
\usepackage{epsfig}
\usepackage{cite}
\usepackage{axodraw}
\usepackage{color}
\definecolor{GreenYellow}   {cmyk}{0.15,0,0.69,0}
\definecolor{Yellow}        {cmyk}{0,0,1,0}
\definecolor{Goldenrod}     {cmyk}{0,0.10,0.84,0}
\definecolor{Dandelion}     {cmyk}{0,0.29,0.84,0}
\definecolor{Apricot}       {cmyk}{0,0.32,0.52,0}
\definecolor{Peach}         {cmyk}{0,0.50,0.70,0}
\definecolor{Melon}         {cmyk}{0,0.46,0.50,0}
\definecolor{YellowOrange}  {cmyk}{0,0.42,1,0}
\definecolor{Orange}        {cmyk}{0,0.61,0.87,0}
\definecolor{BurntOrange}   {cmyk}{0,0.51,1,0}
\definecolor{Bittersweet}   {cmyk}{0,0.75,1,0.24}
\definecolor{RedOrange}     {cmyk}{0,0.77,0.87,0}
\definecolor{Mahogany}      {cmyk}{0,0.85,0.87,0.35}
\definecolor{Maroon}        {cmyk}{0,0.87,0.68,0.32}
\definecolor{BrickRed}      {cmyk}{0,0.89,0.94,0.28}
\definecolor{Red}           {cmyk}{0,1,1,0}
\definecolor{OrangeRed}     {cmyk}{0,1,0.50,0}
\definecolor{RubineRed}     {cmyk}{0,1,0.13,0}
\definecolor{WildStrawberry}{cmyk}{0,0.96,0.39,0}
\definecolor{Salmon}        {cmyk}{0,0.53,0.38,0}
\definecolor{CarnationPink} {cmyk}{0,0.63,0,0}
\definecolor{Magenta}       {cmyk}{0,1,0,0}
\definecolor{VioletRed}     {cmyk}{0,0.81,0,0}
\definecolor{Rhodamine}     {cmyk}{0,0.82,0,0}
\definecolor{Mulberry}      {cmyk}{0.34,0.90,0,0.02}
\definecolor{RedViolet}     {cmyk}{0.07,0.90,0,0.34}
\definecolor{Fuchsia}       {cmyk}{0.47,0.91,0,0.08}
\definecolor{Lavender}      {cmyk}{0,0.48,0,0}
\definecolor{Thistle}       {cmyk}{0.12,0.59,0,0}
\definecolor{Orchid}        {cmyk}{0.32,0.64,0,0}
\definecolor{DarkOrchid}    {cmyk}{0.40,0.80,0.20,0}
\definecolor{Purple}        {cmyk}{0.45,0.86,0,0}
\definecolor{Plum}          {cmyk}{0.50,1,0,0}
\definecolor{Violet}        {cmyk}{0.79,0.88,0,0}
\definecolor{RoyalPurple}   {cmyk}{0.75,0.90,0,0}
\definecolor{BlueViolet}    {cmyk}{0.86,0.91,0,0.04}
\definecolor{Periwinkle}    {cmyk}{0.57,0.55,0,0}
\definecolor{CadetBlue}     {cmyk}{0.62,0.57,0.23,0}
\definecolor{CornflowerBlue}{cmyk}{0.65,0.13,0,0}
\definecolor{MidnightBlue}  {cmyk}{0.98,0.13,0,0.43}
\definecolor{NavyBlue}      {cmyk}{0.94,0.54,0,0}
\definecolor{RoyalBlue}     {cmyk}{1,0.50,0,0}
\definecolor{Blue}          {cmyk}{1,1,0,0}
\definecolor{Cerulean}      {cmyk}{0.94,0.11,0,0}
\definecolor{Cyan}          {cmyk}{1,0,0,0}
\definecolor{ProcessBlue}   {cmyk}{0.96,0,0,0}
\definecolor{SkyBlue}       {cmyk}{0.62,0,0.12,0}
\definecolor{Turquoise}     {cmyk}{0.85,0,0.20,0}
\definecolor{TealBlue}      {cmyk}{0.86,0,0.34,0.02}
\definecolor{Aquamarine}    {cmyk}{0.82,0,0.30,0}
\definecolor{BlueGreen}     {cmyk}{0.85,0,0.33,0}
\definecolor{Emerald}       {cmyk}{1,0,0.50,0}
\definecolor{JungleGreen}   {cmyk}{0.99,0,0.52,0}
\definecolor{SeaGreen}      {cmyk}{0.69,0,0.50,0}
\definecolor{Green}         {cmyk}{1,0,1,0}
\definecolor{ForestGreen}   {cmyk}{0.91,0,0.88,0.12}
\definecolor{PineGreen}     {cmyk}{0.92,0,0.59,0.25}
\definecolor{LimeGreen}     {cmyk}{0.50,0,1,0}
\definecolor{YellowGreen}   {cmyk}{0.44,0,0.74,0}
\definecolor{SpringGreen}   {cmyk}{0.26,0,0.76,0}
\definecolor{OliveGreen}    {cmyk}{0.64,0,0.95,0.40}
\definecolor{RawSienna}     {cmyk}{0,0.72,1,0.45}
\definecolor{Sepia}         {cmyk}{0,0.83,1,0.70}
\definecolor{Brown}         {cmyk}{0,0.81,1,0.60}
\definecolor{Tan}           {cmyk}{0.14,0.42,0.56,0}
\definecolor{Gray}          {cmyk}{0,0,0,0.50}
\definecolor{Black}         {cmyk}{0,0,0,1}
\definecolor{White}         {cmyk}{0,0,0,0}

\def\theequation{\arabic{section}.\arabic{equation}}

\begin{document}
\tolerance=100000

\begin{flushright}
CERN-PH-TH/2009-219, FTUV-09-1118, MAN/HEP/2009/43\\[-1mm]
{\tt arXiv:0911.3611 [hep-ph]}\\[-1mm]
November 2009
\end{flushright}

\bigskip

\begin{center}
{\Large {\bf Flavour Geometry and Effective Yukawa Couplings}}\\[3mm] 
{\Large {\bf         in the MSSM}}\\[1.2cm]
{\large John Ellis}$^{\, a}$,
{\large Robert N. Hodgkinson}$^{\, b,c,e}$,
{\large Jae Sik Lee}$^{\, d}$ and 
{\large Apostolos Pilaftsis}$^{\, e}$\\[0.5cm]
{\it $^a$Theory Division, Physics Department, CERN, CH-1211 Geneva 23,
Switzerland}\\[2mm]
{\it $^b$Theory Group, Institute of Particle and Nuclear Studies,
  KEK,\\ 1-1 Oho Tsukuba-shi, Ibaraki-ken 305-0801, Japan}\\[2mm]
{\it $^c$Departamento de F\'\i sica Te\`orica and IFIC,}\\ 
{\it Universitat de Val\`encia-CSIC, E-46100, Burjassot, 
Val\`encia, Spain}\\[2mm]
{\it $^d$Physics Division, National Centre for Theoretical Sciences,
Hsinchu, Taiwan 300 }\\[2mm]
{\it $^e$School of Physics and Astronomy, University of Manchester}\\
{\it Manchester M13 9PL, United Kingdom}
\end{center}

\vspace*{0.8cm}\centerline{\bf  ABSTRACT}
  \noindent
We present a new geometric approach to the flavour decomposition of an
arbitrary  soft  supersymmetry-breaking   sector  in  the  MSSM.   Our
approach  is based on  the geometry  that results  from the  quark and
lepton Yukawa  couplings, and enables  us to derive the  necessary and
sufficient  conditions for a linearly-independent basis of matrices
related  to  the completeness  of the  internal
$[SU(3)  \otimes  U(1)]^5$  flavour  space.   In  a  second  step,  we
calculate the  effective Yukawa couplings  that are enhanced  at large
values  of $\tan\beta$  for general  soft  supersymmetry-breaking mass
parameters. We highlight the  contributions due to non-universal terms
in  the  flavour decompositions  of  the  sfermion  mass matrices.  We
present numerical examples illustrating  how such terms are induced by
renormalization-group evolution starting from universal input boundary
conditions, and demonstrate their importance for the flavour-violating
effective Yukawa couplings of quarks.
 
\medskip


\vspace*{\fill}
\newpage

\setcounter{equation}{0}
\section{Introduction}
  \label{sec:intro}

Supersymmetry (SUSY) is very attractive as a possible extension of the
Standard  Model,  since it  offers  a  mechanism  for stabilizing  the
hierarchy between  the gravitational and electroweak  scale, would aid
unification of  the gauge couplings, provides a  natural candidate for
the astrophysical  cold dark matter, predicts that  the lightest Higgs
boson  should   be  relatively  light,  could   explain  the  apparent
discrepancy between  the experimental measurement  of $g_\mu -  2$ and
the prediction of  the Standard Model, and is  an essential ingredient
of string theory  \cite{Chung:2003fi}.  However, the mechanism whereby
SUSY  could be  broken is  an open  theoretical question  and,  in the
absence of a convincing  solution, the phenomenological description of
SUSY   breaking  requires  many   unknown  and   apparently  arbitrary
parameters.  These parameters are severely restricted by the
conspicuous success of the Cabibbo-Kobayashi-Maskawa (CKM) description
of flavour violation within the Standard Model\cite{CKM-C,CKM-KM}. Any
model for  new physics at the  TeV scale, such as  SUSY, must maintain
this phenomenological success of the  CKM mixing paradigm: this is the
flavour problem of SUSY.

In this paper we discuss issues related to the SUSY flavour problem in
the context of  the minimal SUSY extension of  the Standard Model, the
MSSM, with  general soft SUSY-breaking parameters.  As  is well known,
these comprise  the U(1)$_Y$, SU(2)$_L$ and  SU(3)$_c$ gaugino masses,
$3  \times 3$ mass-squared  matrices for  the left-handed  squarks and
sleptons and  for the right-handed squarks and  charged sleptons, soft
SUSY-breaking trilinear parameters corresponding to each of the Yukawa
couplings of the Standard Model, soft SUSY-breaking masses for each of
the  two Higgs doublets  $H_{u,d}$ and  a soft  bilinear SUSY-breaking
parameter coupling the Higgs doublets.

The SUSY  flavour problem concerns  the structure of  the mass-squared
matrices  for the  squarks and  sleptons.  In  general,  loop diagrams
involving   sparticle   exchanges   will  contribute   to   low-energy
flavour-changing neutral-current  (FCNC) interactions and CP-violating
observables.   Therefore,  the  agreement  of flavour  data  with  the
predictions    of     the    Standard    Model,     based    on    its
Glashow-Iliopoulos-Maiani mechanism  for suppressing FCNC interactions
\cite{GIM}  and  the   Kobayashi-Maskawa  mechanism  for  CP violation
\cite{CKM-KM},  imposes  stringent  conditions  on the  $3  \times  3$
mass-squared matrices  for the  squarks and sleptons  and on  the soft
trilinear parameters.  It  is often assumed that all  the SUSY flavour
violation is due  to the effects of the  Yukawa coupling matrices, the
minimal              flavour              violation              (MFV)
hypothesis~\cite{Chivukula:1987py,MFV,CNS}.   One way  to  ensure that
this hypothesis is  obeyed is to require each of the mass-squared
matrices  and   the  trilinear  parameters  to   be  universal,  i.e.,
proportional to the unit matrix in flavour space, ${\bf 1}_3$, at some
high  input  scale  prior  to  their  renormalization  by  the  Yukawa
couplings ${\bf h}_{u,d,e}$, which we denote here for brevity as ${\bf
  h}$. However, this is not the most general possibility, and 
  in~\cite{MCPMFV} the maximally CP-violating version of this MFV
model  (MCPMFV) has  been introduced,  which has  19  free parameters,
including 6 CP-violating phases.

As    pointed   out   in~\cite{MCPMFV},  generalizations  that satisfy   the   MFV
hypothesis may  include non-universal  terms  in the  mass-squared
matrices  $\widetilde{\bf M}^2_{Q,L,D,U,E}$  that are  proportional to
Hermitian products of Yukawa couplings ${\bf h} {\bf h}$ and algebraic
functions  of  them. The  renormalization-group  equations (RGEs)  and
their threshold corrections generate  non-universal terms of this type
\cite{SoftMassRGEs,Grzadkowski:1987wr},   which   are in
the  MFV  spirit  and may  be  consistent  with  the
measurements  of  flavour-violating   neutral  interactions  if  their
coefficients are not too large.

As we discuss in this paper, {\it all} deviations from universality in
the  mass-squared matrices  $\widetilde{\bf  M}^2_{Q,L,D,U,E}$ may  be
expressed in  terms of  the quadratic products  ${\bf h} {\bf  h}$ and
their  quadratic and  cubic combinations.   They  constitute complete,
linearly-independent bases for  the Hermitian matrices $\widetilde{\bf
  M}^2_{Q,L,D,U,E}$,  and  hence   provide  a  convenient  geometrical
framework  for  discussing the  SUSY  flavour  problem.  For  example,
constraints  on  MFV  scenarios   within  the  MSSM  may  usefully  be
formulated as  numerical bounds on  the coefficients of  expansions of
the  mass-squared matrices  in these  Hermitian bases.  In  a specific
numerical  example, we  demonstrate how  non-universal terms  in these
flavour decompositions are  induced by renormalization-group evolution
starting from universal input conditions at a Grand Unification scale.

This formulation of SUSY flavour  geometry can be used to simplify and
systematize  the  calculations  of  a  number of  aspects  of  flavour
violation in SUSY models.  In  this paper we discuss one specific such
application, namely to effective quark Yukawa couplings at large $\tan
\beta$ \cite{Banks:1987iu,Hempfling:1993kv,Carena:1999py}.  We present
the complete  sets of one-loop  SUSY corrections to  the self-energies
for  the  down-  and  up-quark and  charged-lepton  Yukawa  couplings,
highlighting  the  importance of  the  contributions of  non-universal
terms in  the flavour  decompositions of the  squark and  slepton mass
matrices.  We then  use the  same numerical  example as  previously to
demonstrate  the  importance  of  these  terms  for  effective  Yukawa
couplings in the MSSM.

The structure of this paper is as follows. In Section 2 we demonstrate
that the quadratic products of  Yukawa couplings ${\bf h} {\bf h}$ and
their quadratic  and cubic combinations constitute a  basis in flavour
space for  soft SUSY-breaking contributions to the  squark and slepton
mass-squared  matrices. Then, in  Section 3  we discuss  the effective
quark  Yukawa  couplings  at  large  $\tan \beta$  and  the  numerical
importance  of   flavour-non-diagonal  terms  in   these  mass-squared
matrices.  Finally,  the main  results  of  our paper  are  summarised  in
Section~4.

\setcounter{equation}{0} 
\section{The Flavour Geometry of the MSSM}

In this section we discuss  several new aspects of flavour geometry in
the MSSM.  ~First,  we describe our geometric approach  to the flavour
decomposition  of a  general  soft SUSY-breaking  sector  in terms  of
Yukawa  couplings.   We  then  illustrate  within  a  specific  MCPMFV
scenario how  this approach  can be used  to study quantitatively the
flavour structure  of renormalization-group  (RG) effects on  the soft
SUSY-breaking matrices.

\subsection{The Geometric Approach to Flavour}
  \label{sec:MFV}

To set the stage, we first introduce the superpotential of the MSSM:
\begin{equation}
  \label{Wpot}
W_{\rm MSSM}\ =\ \widehat{U}^C {\bf h}_u \widehat{Q} \widehat{H}_u\:
+\:   \widehat{D}^C {\bf h}_d \widehat{H}_d \widehat{Q}  \: +\: 
\widehat{E}^C {\bf h}_e \widehat{H}_d \widehat{L} \: +\: \mu
\widehat{H}_u \widehat{H}_d\ ,
\end{equation}
where $\widehat{H}_{u,d}$  are the  two Higgs chiral  superfields, and
$\widehat{Q}$,  $\widehat{L}$,  $\widehat{U}^C$,  $\widehat{D}^C$  and
$\widehat{E}^C$ are the left-  and right-handed superfields related to
up- and  down-type quarks and  charged leptons.  The  Yukawa couplings
${\bf  h}_{u,d,e}$ are  $3\times  3$ complex  matrices describing  the
charged-lepton and quark masses  and their mixings. Finally, the $\mu$
parameter   in~(\ref{Wpot})  describes   the  mixing   of   the  Higgs
supermultiplets; it has  to be of the electroweak  order for a natural
realization of the Higgs mechanism.

The  required breaking  of SUSY  in nature  is described  by  the
effective soft SUSY-breaking Lagrangian
\begin{eqnarray} 
  \label{Lsoft}
-{\cal L}_{\rm soft} &=& \frac{1}{2}\, \Big(\, M_1\,
 \widetilde{B}\widetilde{B}\: +\: M_2\, \widetilde{W}^i\widetilde{W}^i\: 
+\: M_3\, \tilde{g}^a\tilde{g}^a\,  \ +\ {\rm H.c.}\Big)\: +\:
 \widetilde{Q}^\dagger 
\widetilde{\bf M}^2_Q \widetilde{Q}\: +\:  \widetilde{L}^\dagger
\widetilde{\bf M}^2_L \widetilde{L}\nonumber\\
&&+\: \widetilde{U}^\dagger
\widetilde{\bf M}^2_U \widetilde{U}\: 
+\: \widetilde{D}^\dagger \widetilde{\bf M}^2_D \widetilde{D}\: 
+\: \widetilde{E}^\dagger \widetilde{\bf M}^2_E \widetilde{E}\:
+\: M^2_{H_u} H^\dagger_u H_u\: +\: M^2_{H_d} H^\dagger_d H_d\\
&&+\: \Big(B\mu\, H_u H_d\ +\ {\rm H.c.}\Big)\:
+\: \Big(  \widetilde{U}^\dagger {\bf a}_u \widetilde{Q} H_u  \:
+\: \widetilde{D}^\dagger {\bf a}_d H_d \widetilde{Q}  \: +\: 
\widetilde{E}^\dagger {\bf a}_e H_d \widetilde{L} \ +\ {\rm
 H.c.}\Big)\;.\nonumber 
\end{eqnarray}
In the above, $M_{1,2,3}$ are the soft SUSY-breaking masses associated
with the U(1)$_Y$, SU(2)$_L$  and SU(3)$_c$ gauginos, respectively. In
addition, $M^2_{H_{u,d}}$  and $B\mu$ are  the soft masses  related to
the  Higgs doublets  $H_{u,d}$  and their  bilinear mixing.   Finally,
$\widetilde{\bf   M}^2_{Q,L,D,U,E}$   are   the   $3\times   3$   soft
mass-squared matrices  of squarks and sleptons,  and ${\bf a}_{u,d,e}$
are  the  corresponding  $3\times  3$ soft  Yukawa  coupling  matrices
related to quark and  lepton masses~\footnote{The soft Yukawa coupling
  matrices  ${\bf  a}_{u,d,e}$ may  alternatively  be  defined by  the
  relation:   $({\bf   a}_{u,d,e})_{ij}   =  ({\bf   h}_{u,d,e}   {\bf
    A}_{u,d,e})_{ij}$,    where    the    matrix    elements    $({\bf
    A}_{u,d,e})_{ij}$  are  typically   of  order  $M_{\rm  SUSY}$  in
  gravity-mediated  SUSY-breaking models.  In this  paper we  will use
  both definitions for the  soft SUSY-breaking Yukawa couplings.}.  In
addition  to  the  $\mu$  term, the  unconstrained  CP-violating  MSSM
contains 109 mass parameters.

In order to  study the flavour structure of the  MSSM, we first notice
that  under the  unitary flavour  rotations  of the  quark and  lepton
superfields,
\begin{equation}
  \label{SUPERrot}
\widehat{Q}'\ =\ {\bf U}_Q\, \widehat{Q}\, ,\quad  
\widehat{L}'\ =\ {\bf U}_L\, \widehat{L}\, ,\quad
\widehat{U}'^C\ =\ {\bf U}^*_U\, \widehat{U}^C\, ,\quad  
\widehat{D}'^C\ =\ {\bf U}^*_D\, \widehat{D}^C\, ,\quad
\widehat{E}'^C\ =\ {\bf U}^*_E\, \widehat{E}^C\; ,  
\end{equation}
the complete MSSM Lagrangian  of the theory remains invariant provided
the model parameters are redefined as follows:
\begin{eqnarray}
  \label{RGrot}
{\bf h}_{u,d} &\to & {\bf U}_{U,D}^\dagger\, {\bf h}_{u,d}\, 
{\bf  U}_Q\,,\qquad {\bf h}_e\, \ \to\, \ 
{\bf U}_E^\dagger\, {\bf h}_e\, {\bf  U}_L\,,\nonumber\\
\widetilde{\bf M}^2_{Q,L,U,D,E}  &\to & {\bf  U}_{Q,L,U,D,E}^\dagger\, 
\widetilde{\bf  M}^2_{Q,L,U,D,E}\, {\bf  U}_{Q,L,U,D,E}\,,\nonumber\\
{\bf a}_{u,d} &\to & {\bf U}_{U,D}^\dagger\, {\bf a}_{u,d}\, {\bf
  U}_Q\,,\qquad
{\bf a}_e \,\ \to \,\ {\bf U}_E^\dagger\, {\bf a}_e\, {\bf U}_L\; . 
\end{eqnarray}
The  remaining mass  scales, $\mu$,  $M_{1,2,3}$,  $M^2_{H_{u,d}}$ and
$B\mu$,    do    not    transform    under   the    unitary    flavour
rotations~(\ref{SUPERrot}).   Thus,  in  the  absence  of  the  Yukawa
couplings and  soft SUSY-breaking  parameters, the MSSM  possesses the
flavour symmetry $[SU(3)\times U(1)]^5$~\cite{Chivukula:1987py}.

Given the flavour transformations~(\ref{RGrot}), one may wonder whether
the    soft    SUSY-breaking    mass    parameters,    $\widetilde{\bf
  M}^2_{Q,L,U,D,E}$ and  ${\bf a}_{u,d,e}$, can be  expressed in terms
of the  Yukawa couplings ${\bf  h}_{u,d,e}$.  This question  was first
discussed   in~\cite{MCPMFV}  and   subsequently  studied   in  detail
in~\cite{CNS}.   However, our  approach  and results  differ from  the
algebraic method presented in~\cite{CNS},  which makes explicit use of
the  Cayley--Hamilton  identities.  Instead,  our  method is  entirely
geometric and  so enables us  to derive the necessary  and sufficient
conditions on the completeness of the flavour space.

To  start with,  let us  first  consider the  left-handed squark  mass
matrix  $\widetilde{\bf M}^2_Q$.   For a  given  renormalization scale
$M_X$,  $\widetilde{\bf  M}^2_Q$ may  be  entirely  determined by  the
decomposition:
\begin{equation}
  \label{MQMX}
\widetilde{\bf M}^2_Q (M_X)\ =\ 
\sum\limits_{I=0}^8\;\widetilde{m}^{2,I}_Q(M_X)\; {\bf H}^Q_I(M_X)\; ,
\end{equation}
where the ${\bf H}^Q_I$ are the following $3\times 3$ Hermitian matrices
constructed out of the Yukawa coupling matrices:
\begin{eqnarray}
  \label{HQI}
\Big\{\, {\bf H}^Q_I\, \Big\} \!\!\!& = &\!\!\! \Big\{\, {\bf 1}_3\,,\ 
{\bf h}^\dagger_u {\bf h}_u\,,\  {\bf h}^\dagger_d {\bf h}_d\,,\ 
({\bf  h}^\dagger_u {\bf h}_u)^2\,,\  ({\bf h}^\dagger_d {\bf
  h}_d)^2\,,\ \big[\, {\bf h}^\dagger_u {\bf h}_u\,, {\bf h}^\dagger_d {\bf
  h}_d\,\big]_+\,,\  i\big[\, {\bf h}^\dagger_u {\bf
  h}_u\,, {\bf h}^\dagger_d {\bf h}_d\,\big]_-\,,\ \nonumber\\
\!\!\!&&\!\!\! ~~{\bf  h}^\dagger_u {\bf h}_u {\bf h}^\dagger_d {\bf
  h}_d {\bf  h}^\dagger_u {\bf h}_u\,,\ 
  {\bf h}^\dagger_d {\bf
  h}_d {\bf  h}^\dagger_u {\bf h}_u {\bf h}^\dagger_d {\bf
  h}_d\,\Big\}\;.
\end{eqnarray}
In the above,  the index $I$ labels all the  9 matrices ${\bf H}^Q_I$,
i.e.~$I=0,1,2,\dots,8$,  and $[A\,,B]_{\pm}  =  \frac{1}{2}\, (AB  \pm
BA)$  for  two matrices  $A$  and  $B$.   The mass-squared  parameters
$\widetilde{m}^{2,I}_Q(M_X)$  are  all  real  and  parametrize  the  9
independent  elements  of  the   $3\times  3$  Hermitian  mass  matrix
$\widetilde{\bf M}^2_Q (M_X)$.  Hence,  the matrices ${\bf H}^Q_I$ may
be  regarded  as a  complete  and  linearly-independent  set of  basis
vectors (or  matrices) in this  9-dimensional space. The  selection of
the flavour basis  is not unique, but the  choice made in~(\ref{HQI})
is  minimal in  terms of  the number  of the  Yukawa-coupling matrices
${\bf  h}_{u,d}$ involved, and is symmetric under  the exchange  of ${\bf
  h}_u$  with  ${\bf h}_d$,  with the exception  ${\bf  H}^Q_6$, which  is
anti-symmetric.

The flavour  space spanned by ${\bf  H}^Q_I$ may be  assigned a metric
defined by
\begin{equation}
  \label{gQ}
g^Q_{IJ}\ =\ {\rm Tr}\, ( {\bf H}^Q_I\, {\bf H}^Q_J)\; .
\end{equation}
Using the  basis~(\ref{HQI}), we have checked that  the determinant of
the  $9\times  9$-dimensional   matrix  $g^Q_{IJ}$  does  not  vanish,
i.e.~${\rm   det}\,  (g^Q_{IJ})   \neq  0$,   provided   the  Jarlskog
determinant~\cite{Jarlskog} is not zero,
\begin{equation}
  \label{HQcond}
{\rm det}\, {\bf H}^Q_6\ =\ {\rm det}\,\Big(i\big[\, {\bf h}^\dagger_u
{\bf h}_u\,, {\bf h}^\dagger_d {\bf h}_d\,\big]_-\Big)\ \neq\ 0\;.
\end{equation}
The  latter is  a necessary  and sufficient  condition for  the metric
$g^Q_{IJ}$ to be non-degenerate, and  hence for the basis matrices $\{
{\bf H}^Q_I \}$ defined  in~(\ref{HQI}) to form a linearly-independent
set.  This  means that any  arbitrary form of  $\widetilde{\bf M}^2_Q$
can always be expressed in  terms of the basis matrices ${\bf H}^Q_I$.
Given the soft squark  mass-squared matrix $\widetilde{\bf M}^2_Q$ and
the basis  vectors ${\bf H}^Q_I$,  one can project out  the parameters
$\widetilde{m}^{2,I}_Q$ as follows:
\begin{equation}
  \label{m2Q}
\widetilde{m}^{2,I}_Q\ =\ g^{Q,IJ}\;  
{\rm Tr}\, ( {\bf H}^Q_J\, \widetilde{\bf M}^2_Q)\; ,
\end{equation}
where summation over repeated  indices is understood and $g^{Q,IJ}$ is
the   inverse    metric   of   $g^Q_{IJ}$,    obeying   the   property
$g^{Q,IK}g^Q_{KJ} = \delta^I_{\ J}$.

Under  an  unitary  flavour  rotation  (\ref{SUPERrot})  of  the
left-handed  quark  superfields,  the  basis  matrices  ${\bf  H}^Q_I$
transform as follows:
\begin{eqnarray}
  \label{HQtrans}
{\bf H}^Q_I\ \to\ {\bf H}'^Q_I\ \equiv\ {\bf  U}^\dagger_Q\, {\bf H}^Q_I\,
{\bf U}_Q\ =\ (L^Q)_I^{\, J}\, {\bf H}^Q_J\; ,
\end{eqnarray}
where $(L^Q)_I^{\,  J}$ is  a $9  \times 9$ real  matrix which  can be
evaluated from
\begin{equation}
  \label{LQtrans}
(L^Q)_I^{\, J}\ =\ g^{Q, JK}\, {\rm Tr}\, ( {\bf U}^\dagger_Q\, {\bf H}^Q_I\,
{\bf U}_Q\, {\bf H}^Q_K)\; .
\end{equation}
Using~(\ref{HQtrans}),  it is not difficult to  show that the
$9\times  9$  real  transformation  matrix $(L^Q)_I^{\,  J}$  has  the
property:
\begin{equation}
 \label{LQid}
(L^Q)_I^{\, K}\, (L^Q)_J^{\, M}\, g^Q_{KM}\ =\ g^Q_{IJ}\; .
\end{equation}
The  choice of  basis made  in~(\ref{HQI}) is  not orthonormal,
since  $g^Q_{IJ}  \neq \alpha\,  \delta_{IJ}$,  where  $\alpha$ is  an
overall normalization  constant. Had  we chosen the  orthonormal basis
spanned  by the  well-known SU(3)  Gell-Man matrices,  ${\bf  H}^Q_I =
\lambda^I$    (for   $I=1,2,\dots,    8$)   and    ${\bf    H}^Q_0   =
\sqrt{\frac{2}{3}}\;   {\bf   1}_3$,  for   which   $g^Q_{IJ}  =   2\,
\delta_{IJ}$, the matrix $(L^Q)_I^{\, J}$ would have taken the form of
a $9\times 9$ real orthogonal matrix.

One can, by analogy, define basis matrices for the remaining matrices in
the soft SUSY-breaking sector.   Specifically, the $3\times 3$ squared
mass matrix  $\widetilde{\bf M}^2_{U}$ pertaining  to the right-handed
up-type squarks may be decomposed in terms of the basis vectors
\begin{eqnarray}
  \label{HUI}
\Big\{\, {\bf H}^U_I\, \Big\} \!\!\!& = &\!\!\! \Big\{\, {\bf 1}_3\,,\ 
{\bf h}_u {\bf h}^\dagger_u\,,\  
{\bf h}_u {\bf h}^\dagger_d {\bf h}_d {\bf h}^\dagger_u\,, \
({\bf h}_u {\bf h}^\dagger_u)^2\,,\  {\bf h}_u ({\bf h}^\dagger_d {\bf
  h}_d)^2\, {\bf h}^\dagger_u \,,\ 
{\bf h}_u [\, {\bf h}^\dagger_u {\bf h}_u\,, {\bf h}^\dagger_d {\bf
  h}_d\,]_+ {\bf h}^\dagger_u \,,\nonumber\\  
\!\!\!&&\!\!\ 
~~i{\bf h}_u [\, {\bf h}^\dagger_u {\bf h}_u\,, {\bf h}^\dagger_d {\bf
  h}_d\,]_- {\bf h}^\dagger_u\,,\ 
{\bf h}_u {\bf  h}^\dagger_u {\bf h}_u {\bf h}^\dagger_d {\bf
  h}_d {\bf  h}^\dagger_u {\bf h}_u {\bf h}^\dagger_u \,,\
{\bf h}_u {\bf h}^\dagger_d {\bf
  h}_d  {\bf  h}^\dagger_u {\bf h}_u {\bf h}^\dagger_d {\bf
  h}_d {\bf h}^\dagger_u\,\Big\}\;.\qquad
\end{eqnarray}
Again,  we  have  checked  that the  corresponding  metric  $g^U_{IJ}$
defined    as   in~(\ref{gQ})    is   non-degenerate    provided   the
condition~(\ref{HQcond}) holds true and ${\rm det}\,{\bf h}_u \neq 0$.
Likewise,   an   appropriate  basis   $\{   {\bf   H}^D_{I}  \}$   for
$\widetilde{\bf  M}^2_{D}$ may  be  obtained by  replacing the  Yukawa
coupling  matrix ${\bf  h}_d$ with  ${\bf h}_u$  and {\it  vice versa}
in~(\ref{HUI}).  This  basis is non-degenerate  if both~(\ref{HQcond})
and  ${\rm  det}\,{\bf  h}_d  \neq  0$  are  satisfied.  

Finally, the  soft-trilinear Yukawa matrices ${\bf  a}_{u,d}$ may also
be expanded as follows:
\begin{equation}
  \label{aud}
{\bf a}_{u}\ =\ \sum\limits_{I=0}^8\, a^I_{u}\, {\bf h}_{u}\,
{\bf H}^Q_I\;,\qquad
{\bf a}_{d}\ =\ \sum\limits_{I=0}^8\, a^I_{d}\, {\bf h}_{d}\,
{\bf H}^Q_I\; ,
\end{equation}
where  $a^I_{u,d}$ are complex  parameters. Again,  one has  to assume
here that the condition~(\ref{HQcond})  and ${\rm det}\, {\bf h}_{u,d}
\neq  0$ are  satisfied, so  that the  sets $\{  {\bf  h}_{u,d} {\bf
H}^Q_I\}$ form complete bases.

Unlike the scalar  quark sector, the MSSM scalar  lepton sector cannot
be  expanded in  a  complete set  of  basis matrices,  since the  only
available  Yukawa coupling matrix  is ${\bf  h}_e$. However,  if there
exist  right-handed neutrinos that  interact with  left-handed lepton
superfields $\widehat{L}$  via the  Yukawa couplings ${\bf  h}_\nu$, a
complete   basis  can   be  formed   for   describing  $\widetilde{\bf
  M}^2_{L,E}$ and ${\bf a}_e$. In  this case, one needs to perform the
obvious replacements ${\bf h}_d \to {\bf h}_e$ and ${\bf h}_u \to {\bf
  h}_\nu$   in   the   corresponding  quark-basis   matrices   defined
in~(\ref{HQI}) and~(\ref{HUI}).

An interesting  flavour scenario for  the MSSM 
is that termed in~\cite{MCPMFV} the Maximal  CP  and  Minimal  Flavour
Violation (MCPMFV) scenario.  It contains the  following set of
flavour-singlet mass scales at some input scale $M_X$ that may be identical
with $M_{\rm GUT}$:
\begin{equation}
  \label{MCPMFV}
M_{1,2,3}\,,\quad M^2_{H_{u,d}}\,,\qquad 
\widetilde{\bf  M}^2_{Q,L,U,D,E}\ =\ \widetilde{M}^2_{Q,L,U,D,E}\,{\bf
  1}_3\,,\qquad
{\bf A}_{u,d,e}\ =\ A_{u,d,e}\, {\bf 1}_3\; ,
\end{equation}
with  the obvious identifications:  $\widetilde{m}^{2,0}_{Q,L,U,D,E} =
\widetilde{M}^2_{Q,L,U,D,E}$ and $a^0_{u,d,e}  = A_{u,d,e}$. At energy
scales below  $M_X$, RG  effects modify the  flavour structure  of the
soft SUSY-breaking  mass and trilinear  matrices.  Specifically, these
matrices get shifted as follows:
\begin{equation}
  \label{MFVshift}
\mbox{\boldmath $\delta$} \widetilde{\bf  M}^2_{Q,L,U,D,E}\ =\ \widetilde{\bf
  M}^2_{Q,L,U,D,E}\: -\: \widetilde{M}^2_{Q,L,U,D,E}\,
  {\bf 1}_3\,,\qquad 
\mbox{\boldmath $\delta$}{\bf a}_{u,d,e}\ =\ {\bf a}_{u,d,e} \: -\:
{\bf h}_{u,d,e}\, A_{u,d,e}\; ,
\end{equation}
where     $\widetilde{M}^2_{Q,L,U,D,E}     =    \frac{1}{3}\,     {\rm
Tr}\,(\widetilde{\bf    M}^2_{Q,L,U,D,E}   )$    and    $A_{u,d,e}   =
\frac{1}{3}\, {\rm Tr}\,  ({\bf h}^{-1}_{u,d,e} {\bf a}_{u,d,e})$.

In  the next  section we  present numerical  examples  of RG-generated
flavour  structures   in  the  squark  mass   matrices  and  trilinear
couplings.   We  do  not  enter  into numerical  calculations  of  the
coefficients  in a  flavour  decomposition of  the soft  SUSY-breaking
parameters in  the lepton sector, as  these would be  dependent on the
model for the neutrino sector.  In Section~\ref{threshold_corrections}
we then use the MFV basis decomposition developed here and the flavour
shifts    given   in~(\ref{MFVshift})    to   obtain    the   complete
flavour-covariant  structure  of  the  threshold  corrections  to  the
effective Yukawa couplings at leading order.

\subsection{Flavour Geometry of Renormalization Group Effects}

We now present numerical calculations of the flavour structure of
the  soft SUSY-breaking  mass and  trilinear matrices  for  a specific
family of MCPMFV scenarios with
\begin{eqnarray}
&&\left|M_{1,2,3}\right|=250~~{\rm GeV}\,, \nonumber \\
&&M^2_{H_u}=M^2_{H_d}=\widetilde{M}^2_Q=\widetilde{M}^2_U=\widetilde{M}^2_D
=\widetilde{M}^2_L=\widetilde{M}^2_E=(100~~{\rm GeV})^2\,, \nonumber \\
&&\left|A_u\right|=\left|A_d\right|=\left|A_e\right|=100~~{\rm GeV}\,,
\qquad
\Phi_{A}^{\rm GUT}\equiv\Phi_{A_u}=\Phi_{A_u}=\Phi_{A_e}=0^\circ\,,\quad
\label{eq:cpsps1a}
\end{eqnarray}
at the GUT scale, varying the input value of $\tan \beta$ between
10 and 50. When $\tan \beta = 10$, this choice of parameters
corresponds approximately to benchmark point B~\cite{bench} and SPS
point 1a~\cite{SPS}. We also allow for various common values of the
CP-violating gaugino  phases, denoted by $\Phi_M$.

\begin{figure}[p]
\hspace{ 0.0cm}
\centerline{\epsfig{figure=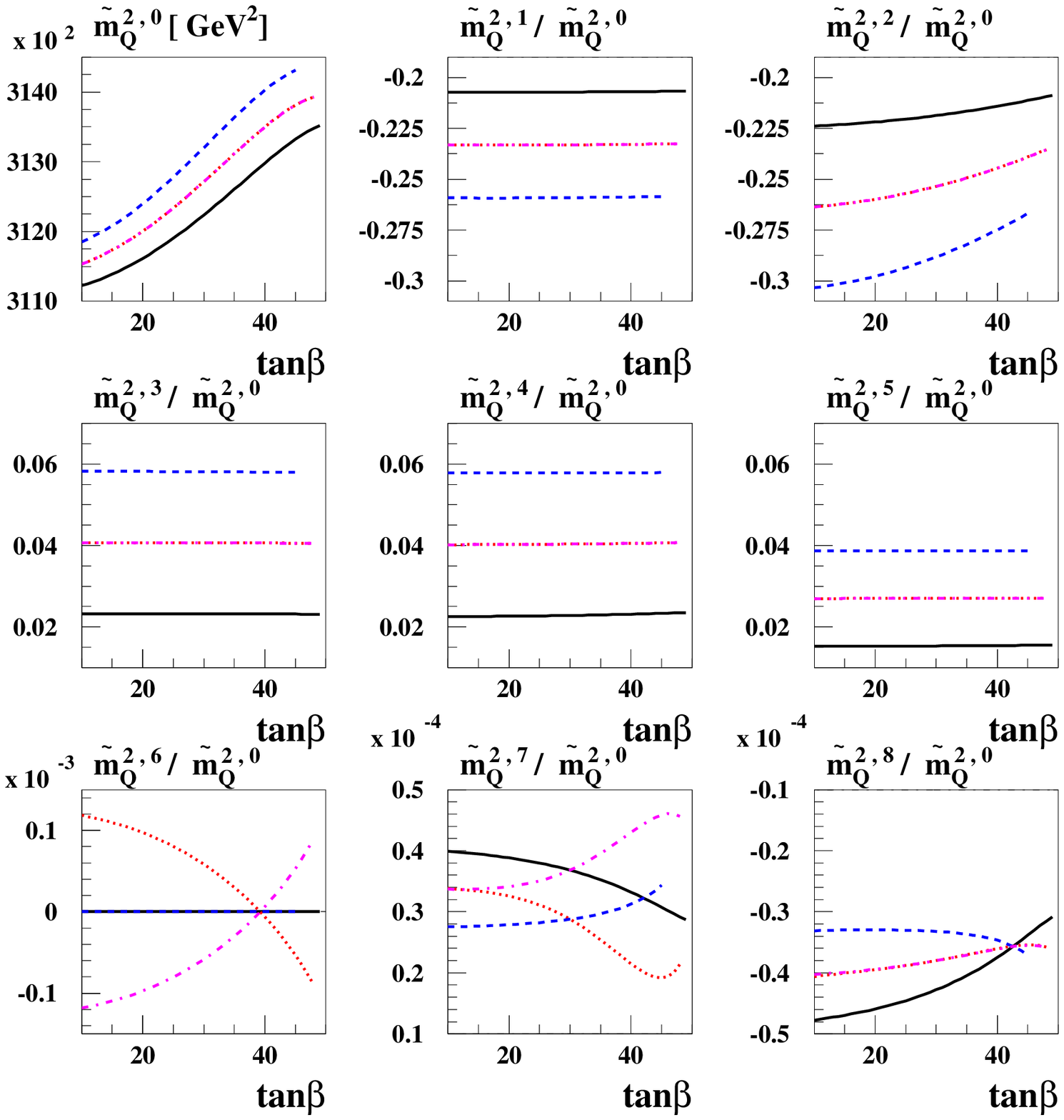,height=16cm,width=16cm}}
\vspace{-0.9cm}
\caption{\it In the top left panel we display the variation of
  $\widetilde{m}^{2,0}_Q$ with $\tan \beta$, in the expansion
  (\protect\ref{MQMX}) of the left-handed squark mass-squared matrix,
  as evaluated at the SUSY-breaking scale.
In the other panels we display the variations with $\tan \beta$ of the ratios
$\widetilde{m}^{2,I}_Q/\widetilde{m}^{2,0}_Q$.
In each frame, the solid (black), dotted (red), dashed (blue), and
dash-dotted (magenta) lines are for 
$\Phi_M\equiv \Phi_1 = \Phi_2 = \Phi_3 = 0^\circ$, $90^\circ$, $180^\circ$, 
and $270^\circ$, respectively.
The input MCPMFV SUSY-breaking parameters are taken as in
(\protect\ref{eq:cpsps1a}).  }
\label{fig:cmq2}
\end{figure}

As  a  first example,  Fig.~\ref{fig:cmq2}  gives some  representative
numerical  results for  this  choice of  parameters,  showing how  the
different  coefficients   $\widetilde{m}^{2,I}_Q(M_X)$  in  the  basis
expansion (\ref{MQMX})  of the left-handed  squark mass-squared matrix
$\widetilde{\bf M}^2_Q  (M_X)$ vary as functions of  $\tan \beta$ when
evaluated at the SUSY-breaking scale, for various values of 
$\Phi_M$.  Note that the  dotted (red) and
dash-dotted  (magenta) lines,  corresponding to  $\Phi_M=90^\circ$ and
$270^\circ$   respectively,  overlap   for  all   coefficients  except
$\widetilde{m}^{2,6}_Q$ and $\widetilde{m}^{2,7}_Q$; the solid (black)
and  dashed   (blue)  lines  corresponding   to  $\Phi_M=0^\circ$  and
$180^\circ$ overlap for $\widetilde{m}^{2,6}_Q$.

Specifically, the  top left panel  of Fig.~\ref{fig:cmq2} demonstrates
how the  leading flavour-singlet piece  $\widetilde{m}^{2,0}_Q$ varies
with $\tan  \beta$: we  see that  the variation is  well below  the \%
level.   Turning to  the  other panels  of Fig.~\ref{fig:cmq2},  which
display the  ratios of the  other coefficients $\widetilde{m}^{2,I}_Q$
to $\widetilde{m}^{2,0}_Q$, we note first that  the coefficients are
all  ${\cal  O}(1)$ or  smaller,  showing  that our  flavour-geometric
expansion  is well  behaved. In  particular, there  are  no unphysical
divergences  for  large $\tan  \beta$.   The largest  flavour-changing
contributions  are  generated  along the  $\widetilde{m}^{2,1}_Q$  and
$\widetilde{m}^{2,2}_Q$   directions,  corresponding   to   the  basis
matrices   with   the   fewest    powers   of   $\bf{hh}$.    Such   a
hierarchical structure amongst the coefficients can be understood
in terms of  an approximate iterative solution to  the RGEs.  Finally,
we observe that the  magnitude of $\widetilde{m}^{2,I}_Q$ is maximised
when $\Phi_M=180^\circ$ for $I=1,5$.

\begin{figure}[tp]
\hspace{ 0.0cm}
\centerline{\epsfig{figure=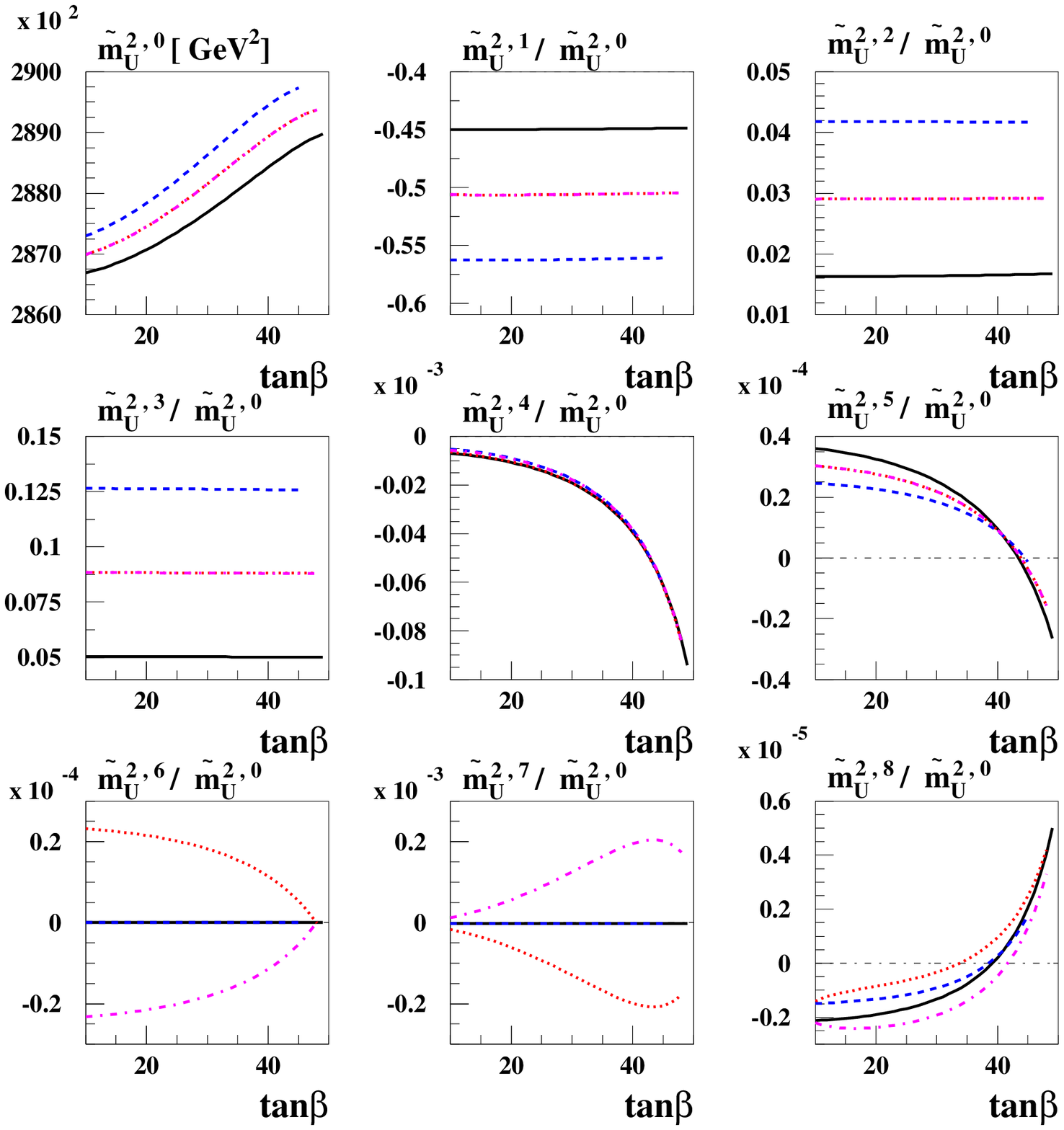,height=16cm,width=16cm}}
\vspace{-0.9cm}
\caption{\it In the top left panel we display the variation of
  $\widetilde{m}^{2,0}_U$ with $\tan \beta$, in the expansion
  corresponding to (\protect\ref{MQMX}) for the right-handed up-squark
  mass matrix, as evaluated at the SUSY-breaking scale. In the other
  panels we display the variations with $\tan \beta$ of the ratios
  $\widetilde{m}^{2,I}_U/\widetilde{m}^{2,0}_U$.  The lines are the
  same as in Fig.~\ref{fig:cmq2}.  The input MCPMFV SUSY-breaking
  parameters are taken as in (\protect\ref{eq:cpsps1a}).  }
\label{fig:cmu2}
\end{figure}

\begin{figure}[tp]
\hspace{ 0.0cm}
\centerline{\epsfig{figure=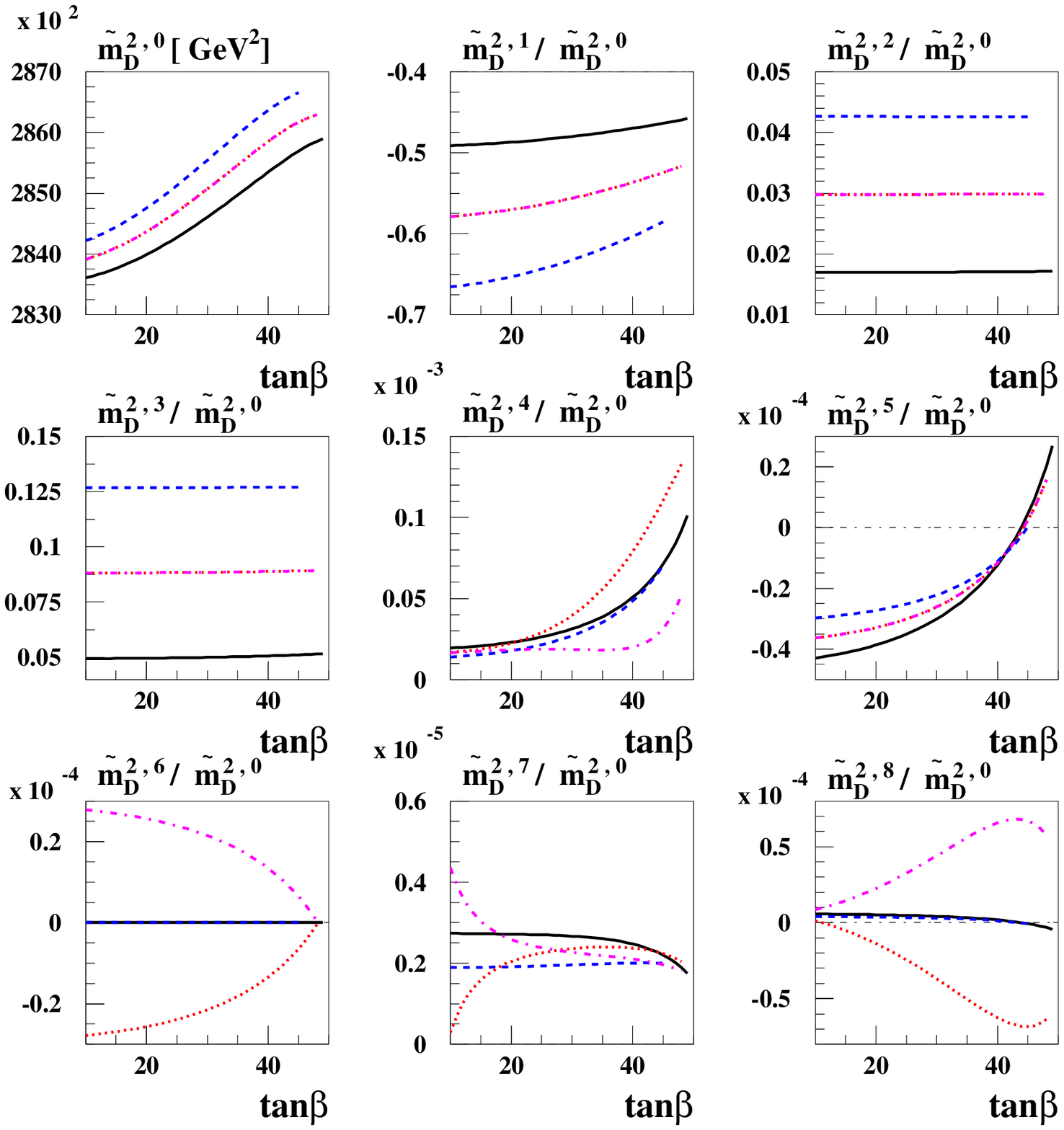,height=16cm,width=16cm}}
\vspace{-0.9cm}
\caption{\it In the top left panel we display the variation of
  $\widetilde{m}^{2,0}_D$ with $\tan \beta$, in the expansion
  corresponding to (\protect\ref{MQMX}) for the right-handed
  down-squark mass matrix, as evaluated at the SUSY-breaking scale. In
  the other panels we display the variations with $\tan \beta$ of the
  ratios $\widetilde{m}^{2,I}_D/\widetilde{m}^{2,0}_D$.  The lines are
  the same as in Fig.~\ref{fig:cmq2}.  The input MCPMFV SUSY-breaking
  parameters are taken as in (\protect\ref{eq:cpsps1a}).  }
\label{fig:cmd2}
\end{figure}

Some remarks  regarding the computational procedure  followed here are
in order.  In our approach, we  fix the $3\times 3$ Yukawa matrices at
the scale $m_t^{\rm pole}$ by applying the boundary conditions
\begin{equation}
{\bf h}_u\left(m_t^{\rm pole}\right) = \frac{\sqrt 2}{v} \widehat{\bf
  M}_u\left(m_t^{\rm pole}\right)\,,\qquad 
{\bf h}_{d,e}\left(m_t^{\rm pole}\right) = \frac{\sqrt 2}{v}
\widehat{\bf M}_{d,e}\left(m_t^{\rm pole}\right) 
{\bf V}_{d,e}^\dag\left(m_t^{\rm pole}\right)\ ,
\end{equation}
where ${\bf V}_d$ is the  physical CKM matrix, ${\bf V}^\dag_e$ is the
Pontecorvo-Maki-Nakagawa-Sakata (PMNS)  matrix~\cite{PMNS},  
and the $\widehat{\bf  M}_f$ are  the  diagonal
fermion  mass matrices.   These  couplings run  to  the scale  $M_{\rm
  SUSY}$ according to the SM RGEs, where they match as
\begin{eqnarray}
{\bf h}_u\left(M_{\rm SUSY}^+\right) & = & {\bf h}_u\left(M_{\rm
  SUSY}^-\right) / \sin\beta\left(M_{\rm SUSY}\right)\ , 
\nonumber\\
{\bf h}_{d,e}\left(M_{\rm SUSY}^+\right) & = & {\bf
  h}_{d,e}\left(M_{\rm SUSY}^-\right) / \cos\beta\left(M_{\rm
  SUSY}\right)\ , 
\end{eqnarray}
before  running from the  soft SUSY-breaking  scale $M_{\rm  SUSY}$ to
$M_{\rm GUT}$ according  to the RGEs of the  MSSM.  However, threshold
corrections  due to  sfermion-gaugino and  sfermion-Higgsino exchange,
discussed  in  detail  in  the  next section,  can  modify  the  above
relations so that we have instead at the soft SUSY-breaking scale,
\begin{eqnarray}
{\bf h}_u\left(M_{\rm SUSY}\right) & = & \frac{\sqrt{2}}{v
  \sin\beta}\, \widehat{\bf M}_u\left(M_{\rm SUSY}\right)\,  
{\bf R}^{-1}_u\left(M_{\rm SUSY}\right)\ ,\nonumber\\
{\bf h}_{d,e}\left(M_{\rm SUSY}\right) & = & \frac{\sqrt{2}}{v
  \cos\beta}\, \widehat{\bf M}_{d,e}\left(M_{\rm SUSY}\right)\,  
{\bf V}_{d,e}^\dagger\left(M_{\rm SUSY}\right)\, {\bf
  R}^{-1}_{d,e}\left(M_{\rm SUSY}\right)\ , 
\end{eqnarray}
where  the  ${\bf  R}_{u,d,e}$  matrices resum  potentially  large  or
$\tan\beta$-enhanced effects due to threshold corrections. Since these
corrections  depend upon  the  soft SUSY-breaking  mass and  trilinear
matrices  evaluated  at  the  SUSY-breaking  scale,  rather  than  the
GUT-scale,  a full implementation  of these  effects would  require an
iterative  solution to  the  RGEs.   Instead, we  follow  here a  more
simplified   approach   and  treat   the   threshold  corrections   as
higher-order  effects  by  neglecting  their contribution  to  the  RG
running.  Such an approximation is generally valid for most choices of
the  SUSY-breaking parameters  (for exceptional  regions  of parameter
space, see~\cite{DP}).

One  may think  of certain  extreme  scenarios in  which the  threshold
corrections lead to dramatic effects in the low-energy theory and some
care  is required in  describing the  flavour-geometry of  such cases.
For instance, one may consider a scenario in which the Yukawa matrices
are real at the GUT  scale.  Such a scenario was previously considered
in~\cite{Abel:1996eb}. Then,  Higgsino-mediated threshold effects will
make  the Yukawa  couplings complex,  and so  a  sizeable CP-violating
phase for the CKM matrix  can be generated, especially at large values
of $\tan\beta$.   In this  situation, the condition  (\ref{HQcond}) is
not satisfied  by the running Yukawa matrices  above the SUSY-breaking
scale, and these  matrices cannot be used as a  complete basis for
projecting  out  the  CP-violating  flavour  structure  of  the  soft
SUSY-breaking  matrices.  Instead,  we  may use  the effective  Yukawa
couplings  that include CP-violating  threshold effects  to decompose
the soft SUSY-breaking  mass matrices as normal.  In  such a scenario,
the observed  CP violation  in the $K$-  and $B$-meson systems  may be
accounted  for predominantly  by the  CP-violating  soft SUSY-breaking
sector.

\begin{figure}[thp]
\hspace{ 0.0cm}
\centerline{\epsfig{figure=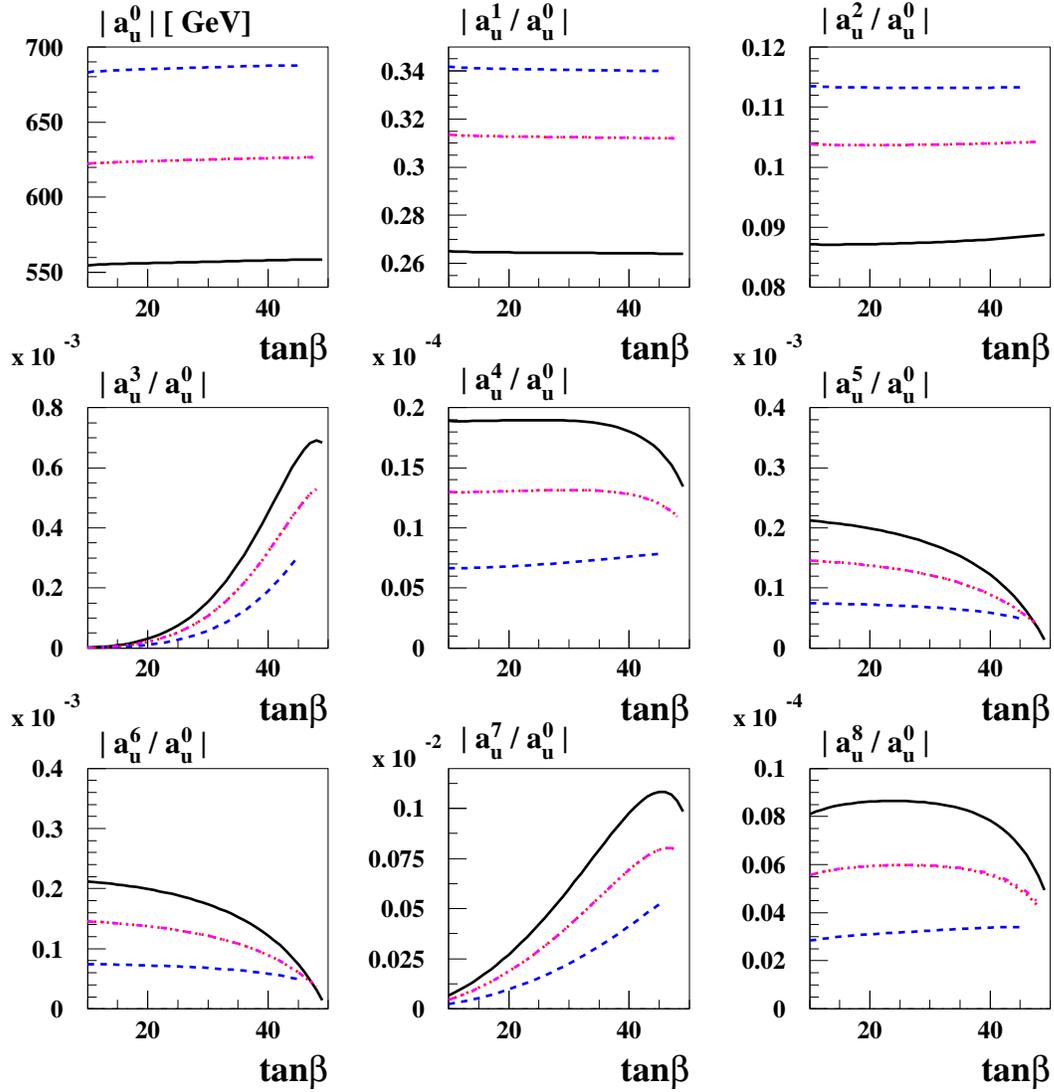,height=16cm,width=16cm}}
\vspace{-0.9cm}
\caption{\it In the top left panel we display the variation of $a^0_u$
  with $\tan \beta$, in the expansion for the trilinear coupling ${\bf
    a}_{u}$ in (\protect\ref{aud}), as evaluated at the SUSY-breaking
  scale. In the other panels we display the variations with $\tan
  \beta$ of the ratios $|a^I_u/a^0_u|$.  The lines are the same as in
  Fig.~\ref{fig:cmq2}.  The input MCPMFV SUSY-breaking parameters are
  taken as in (\protect\ref{eq:cpsps1a}).  }
\label{fig:cau}
\end{figure}

\begin{figure}[thp]
\hspace{ 0.0cm}
\centerline{\epsfig{figure=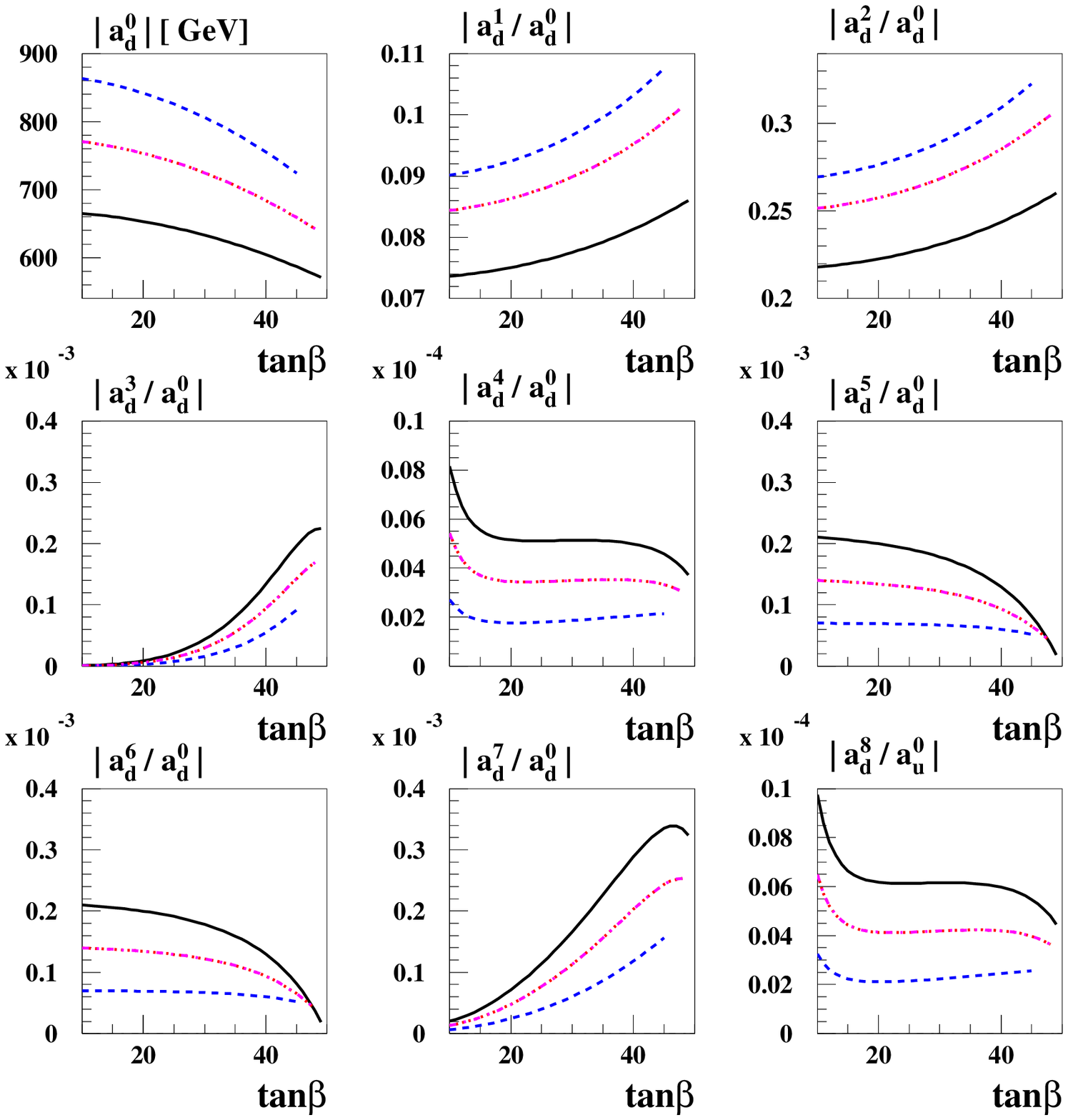,height=16cm,width=16cm}}
\vspace{-0.9cm}
\caption{\it As in Fig.~\protect\ref{fig:cau}, but for variations of
  the coefficients $a^I_d$ with $\tan \beta$, in the expansion for the
  trilinear coupling ${\bf a}_{d}$ in (\protect\ref{aud}).  }
\label{fig:cad}
\end{figure}

Another extremal  case would  be to consider  a scenario in  which the
effective Yukawa  couplings at the  top-quark mass scale are  real, by
arranging  for ${\bf  V}_f^\dagger\, {\bf  R}^{-1}_f$ to  be  real for
$f=d,e$.  In this case, the flavour geometry of the soft SUSY-breaking
matrices  cannot be  decomposed  in  our usual  basis  defined by  the
effective  Yukawa  couplings  and  we  would be  forced  to  choose  a
different basis, e.g.~taking the threshold-uncorrected Yukawa matrices
as defined above the SUSY scale.  In such a scenario the CKM matrix is
complex  at   the  tree  level   and  CP-violation  may   be  mediated
predominantly by the $W$ bosons according to the standard KM paradigm.
In contrast,  the $\tan^2\beta$-enhanced Higgs-mediated  effects would
be purely CP-conserving.

Figure~\ref{fig:cmu2} gives some  representative numerical results for
the   choice  of  parameters   (\ref{eq:cpsps1a}),  showing   how  the
RG-induced    coefficients   in    the   flavour    decomposition   of
$\widetilde{\bf M}^2_{U}$  vary as functions of $\tan  \beta$.  We see
that  the  lines corresponding  to  $\Phi_M=90^\circ$ and  $270^\circ$
again  overlap   for  the  coefficients  $I=0,3$,   whilst  those  for
$\Phi_M=0^\circ$  and  $180^\circ$ are  overlapping  for $I=6,7$.   We
again    observe   a    hierarchical   structure    amongst   the
flavour-changing  components, although  now  with a  bias towards  the
up-type      quark     Yukawa      matrices.       The     coefficient
$\widetilde{m}^{2,1}_U$, corresponding  to the $\bf{h}_u\bf{h}_u^\dag$
direction  in flavour  space,  is here  of  order $\sim  50\%$ of  the
leading  flavour-singlet  term.   In  contrast, the  down-type  Yukawa
couplings  first  enter at  the  order  $({\bf  hh})^2$ and  are  thus
suppressed.

By analogy with Figs.~\ref{fig:cmq2} and \ref{fig:cmu2},
Figs.~\ref{fig:cmd2}, \ref{fig:cau} and \ref{fig:cad} show for the
same choice of parameters (\ref{eq:cpsps1a}) the variations with $\tan
\beta$ of the corresponding coefficients in the flavour decompositions
of the right-handed down-squark mass-squared matrices, and the
trilinear couplings ${\bf a}_{u,d}$, respectively.  We see that the
coefficients of $\widetilde{\bf M}^2_{D}$ exhibit a similar behaviour
to those of $\widetilde{\bf M}^2_{U}$, favouring now the directions in
flavour space corresponding to the down-type quark Yukawa matrices and
consequently with a stronger dependence on $\tan\beta$.  Note also the
relative change in sign between $\widetilde{m}^{2,5}_U$ and
$\widetilde{m}^{2,5}_D$.  Turning to the trilinear couplings, we see a
similar hierarchical pattern amongst the leading terms, with the
largest non-singlet coefficient of $\bf{a}_u(\bf{a}_d)$ being
$I=1~(2)$, corresponding to the
$\bf{h}_u^\dag\bf{h}_u$~$(\bf{h}_d^\dag\bf{h}_d)$ direction in our
$9$-dimensional flavour-space. As before, we observe a stronger
$\tan\beta$-dependence amongst the coefficients of ${\bf a}_d$ than
${\bf a}_u$.  We note that only the $I=0,2$ coefficients of
$\bf{a}_{u,d}$ share the tendency of the sfermion mass matrices to
exhibit the largest RGE effects at $\Phi_M=180^\circ$, with this
behaviour being inverted amongst the remaining coefficients.

\setcounter{equation}{0} 
\section{Effective Yukawa Couplings at Large $\tan\beta$}
\label{threshold_corrections}

In this section  we discuss the one-loop threshold  corrections to the
Yukawa couplings of the MSSM  Higgs bosons to fermions.  We 
include explicitly the contributions  due to  flavour-changing structure  in the
soft  SUSY-breaking terms.  Following  the flavour-covariant  approach
of~\cite{MCPMFV,DP}, we first present the calculation of these effects
in  the  weak basis  for  the  down-type  quarks, up-type  quarks  and
leptons, and then  relate these to the relevant  couplings in the mass
eigenbasis.  Finally, we  present representative numerical results for
the Higgs-boson FCNC couplings that can be relevant for the $B_{d,s}$-
and $K$-meson systems.

\subsection{Threshold Corrections to Yukawa Couplings}

\subsubsection{Down-type Quark Yukawa Couplings}

In  this  section  we  present  the  complete  set  of  one-loop  SUSY
corrections to the  self-energies of the down-type  quarks.  These may
be described by the effective Lagrangian
\begin{equation}
\label{L_eff_down}
-{\mathcal L}^d_{\rm eff}\left[\Phi_1,\Phi_2\right]\
=\
\bar{d}^0_{iR}
\left({\bf h}_d \Phi_1^{\dag\alpha}
+\mbox{\boldmath$\Delta$}{\bf h}_d^\alpha\left[\Phi_1,\Phi_2\right]\right)_{ij}
Q^{0\alpha}_{jL}\ ,
\end{equation}
where $\alpha=1,2$  is a weak-isospin index and  $\Phi_{1(2)}$ are the
scalar components  of the Higgs  doublet superfields giving  masses to
the down-type (up-type) quarks respectively\footnote{Here we adopt the
  convention  for  the  Higgs doublets:  $H_u\equiv\Phi_2,\  H_d\equiv
  i\sigma_2  \Phi_1^\ast$,   where  $\sigma_2$  is   the  usual  Pauli
  matrix.}.  In~(\ref{L_eff_down}) the first term gives the tree-level
contribution,  whilst   $\mbox{\boldmath$\Delta$}  {\bf  h}_d$   is  a
$3\times  3$  matrix  which   is  a  Coleman-Weinberg  type  effective
functional of the background Higgs fields \cite{EffectivePotential},
which we decompose as
\begin{equation}
\mbox{\boldmath$\Delta$}{\bf h}_d\ =\ 
\mbox{\boldmath$\Delta$}{\bf h}_d^{\rm 2HDM}\:
+\: \mbox{\boldmath$\Delta$}{\bf h}_d^{\rm SUSY}\:
+\: \mbox{\boldmath$\Delta$}{\bf h}_d^{\rm CT}\ ,
\end{equation}
where   $\mbox{\boldmath$\Delta$}{\bf  h}_d^{\rm  CT}$   contains  the
counterterms  required to  cancel  the divergences  of  the first  two
terms.   The   contributions  $\mbox{\boldmath$\Delta$}{\bf  h}_d^{\rm
  2HDM}$  are insensitive to  the flavour  structure of  the soft-SUSY
breaking mass and trilinear matrices at the one-loop level and, unless
explicitly  stated, we  neglect  them  in what  follows.   Due to  the
no-renormalization  theorem  for the  SUSY  superpotential, the  MSSM
Yukawa   couplings  are   renormalized  only   by   the  wave-function
counterterms $Z^{1/2}_{\hat D,\hat H_1,\hat Q}$ of the Higgs and quark
superfields, so that
\begin{eqnarray}
\mbox{\boldmath$\Delta$}{\bf h}_d^{\rm CT} & = &
\left(Z^\frac{1}{2}_{\hat D} Z^\frac{1}{2}_{\hat H_1}
Z^\frac{1}{2}_{\hat Q}\: -\: 1\right) {\bf h}_d\,\Phi_1^\dagger\ \simeq\ 
\frac{1}{2} \sum_{i=\hat D,\hat H_1,\hat Q} \delta Z_i\; {\bf h}_d\, 
\Phi_1^\dagger
\end{eqnarray}
with $\delta Z_i  = Z_i - 1$.  Note that there  are no counterterms to
the down-type quark Yukawa coupling proportional to $\Phi_2^\dagger$.

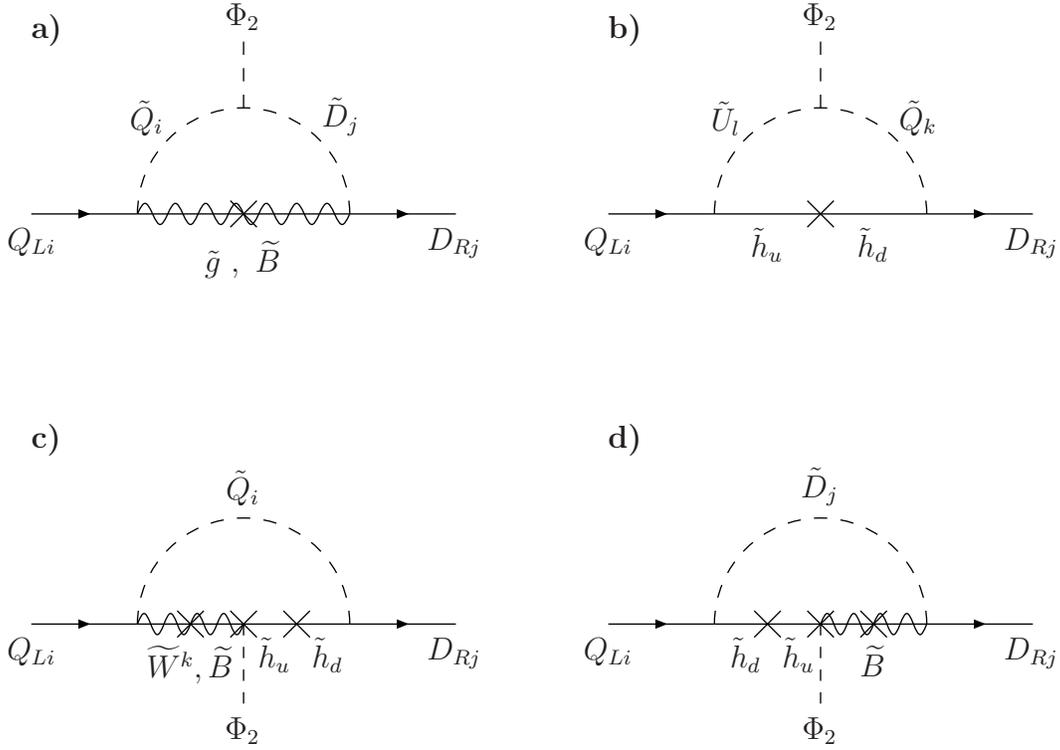
\begin{figure}[t!]
\begin{center}
\begin{tabular}{ccc}
\begin{picture}(180,150)(0,0)
\ArrowLine(10,50)(50,50)
\Line(50,50)(130,50)
\ArrowLine(130,50)(170,50)
\Photon(50,50)(130,50){4}{7}
\DashCArc(90,50)(40,0,180){5}
\DashLine(90,90)(90,115){5}
\Text(90,120)[b]{$\Phi_2$}
\Line(85,45)(95,55)
\Line(95,45)(85,55)
\Text(10,45)[t]{$Q_{Li}$}
\Text(170,45)[t]{$D_{Rj}$}
\Text(60,80)[rb]{$\tilde Q_i$}
\Text(120,80)[lb]{$\tilde D_j$}
\Text(90,40)[t]{$\tilde g\ ,\ \widetilde B$}
\Text(10,120)[l]{\bf a)}
\end{picture}
& \hspace{0.5 cm} &
\begin{picture}(180,150)(0,0)
\ArrowLine(10,50)(50,50)
\Line(50,50)(130,50)
\ArrowLine(130,50)(170,50)
\DashCArc(90,50)(40,0,180){5}
\DashLine(90,90)(90,115){5}
\Text(90,120)[b]{$\Phi_2$}
\Line(85,45)(95,55)
\Line(95,45)(85,55)
\Text(10,45)[t]{$Q_{Li}$}
\Text(170,45)[t]{$D_{Rj}$}
\Text(60,80)[rb]{$\tilde U_l$}
\Text(120,80)[lb]{$\tilde Q_k$}
\Text(70,45)[t]{$\tilde h_u$}
\Text(110,45)[t]{$\tilde h_d$}
\Text(10,120)[l]{\bf b)}
\end{picture}
\\
\begin{picture}(180,150)(0,0)
\ArrowLine(10,50)(50,50)
\Line(50,50)(130,50)
\ArrowLine(130,50)(170,50)
\Photon(50,50)(90,50){4}{4}
\DashCArc(90,50)(40,0,180){5}
\DashLine(90,50)(90,20){5}
\Text(90,15)[t]{$\Phi_2$}
\Line(65,45)(75,55)
\Line(75,45)(65,55)
\Line(85,45)(95,55)
\Line(95,45)(85,55)
\Line(105,45)(115,55)
\Line(115,45)(105,55)
\Text(90,95)[b]{$\tilde Q_{i}$}
\Text(70,42)[t]{$\widetilde W^k,\widetilde B$}
\Text(102,45)[t]{$\tilde h_u$}
\Text(122,45)[t]{$\tilde h_d$}
\Text(10,45)[t]{$Q_{Li}$}
\Text(170,45)[t]{$D_{Rj}$}
\Text(10,120)[l]{\bf c)}
\end{picture}
& \hspace{0.5 cm} &
\begin{picture}(180,150)(0,0)
\ArrowLine(10,50)(50,50)
\Line(50,50)(130,50)
\ArrowLine(130,50)(170,50)
\Photon(90,50)(130,50){4}{4}
\DashCArc(90,50)(40,0,180){5}
\DashLine(90,50)(90,20){5}
\Text(90,15)[t]{$\Phi_2$}
\Line(65,45)(75,55)
\Line(75,45)(65,55)
\Line(85,45)(95,55)
\Line(95,45)(85,55)
\Line(105,45)(115,55)
\Line(115,45)(105,55)
\Text(90,95)[b]{$\tilde D_{j}$}
\Text(110,42)[t]{$\widetilde B$}
\Text(62,45)[t]{$\tilde h_d$}
\Text(82,45)[t]{$\tilde h_u$}
\Text(10,45)[t]{$Q_{Li}$}
\Text(170,45)[t]{$D_{Rj}$}
\Text(10,120)[l]{\bf d)}
\end{picture}
\end{tabular}
\caption{\it The complete set of gauge- and flavour-covariant one-loop
  diagrams contributing to the down-type quark self-energy, to first
  order in $\Phi_2$.  Note that in panels (a) and (c) contributions
  from all listed gauginos should be included.}
\label{CTE_down}
\end{center}
\end{figure}

The  one-loop  SUSY  threshold  corrections  to  the  down-type  quark
self-energy   may  be   calculated  from   the  Feynman   diagrams  of
Fig.~\ref{CTE_down}.  More explicitly, these are given by
\begin{eqnarray}
\label{complete_down}
-\left(\mbox{\boldmath$\Delta$}{\bf h}_d^{\rm SUSY}\right)_{ij}^\alpha
& = & 
\int {d^n k\over(2\pi)^n i}
\left[P_L {-2C_F\,g_3^2 M_3^\ast\over k^2-|M_3^2|}
\left({1\over k^2{\bf 1}_{12}-\widetilde{\bf M}^2}\right)_
{\tilde D_i\tilde Q_j^{\dag\alpha}}
\right.\\
&&
\hspace{-2cm}
+\ P_L \frac{g_1^2}{9}
\left({1\over \not\! k {\bf 1}_8-{\bf M}_C P_L
-{\bf M}_C^\dag P_R}\right)_{\tilde B \tilde B}
P_L
\left({1\over k^2{\bf 1}_{12}-\widetilde{\bf M}^2}\right)_
{\tilde D_i\tilde Q_j^{\dag\alpha}} \nonumber\\
&&
\hspace{-2cm}
+\ P_L\left({1\over \not\! k {\bf 1}_8-{\bf M}_C P_L
-{\bf M}_C^\dag P_R}\right)_{\tilde H_d^\gamma \tilde H_u^\beta}
P_L \left({\bf h}_d\right)_{il}
\left(i\sigma_2\right)^{\gamma\delta}
\left({1\over k^2{\bf 1}_{12}-\widetilde{\bf M}^2}\right)_
{\tilde Q_l^\delta \tilde U_k^\dag} \left({\bf h}_u\right)_{kj}
\left(-i\sigma_2\right)^{\beta\alpha}\nonumber\\
&&
\hspace{-2cm}
+ P_L 
\left({1\over \not\! k{\bf 1}_8
-{\bf M}_C P_L-{\bf M}_C^\dag P_R}\right)_{\tilde H_d^\gamma \tilde B}P_L
\left({\bf h}_d\right)_{il}
\left(i\sigma_2\right)^{\gamma\beta}
\left({1\over k^2{\bf 1}_{12}-\widetilde{\bf M}^2}\right)_{\tilde Q_l^\beta
\tilde Q_j^{\dag\alpha}}\left(\frac{g_1}{3\sqrt{2}}\right)\nonumber\\
&&
\hspace{-2cm}
+\sum_k P_L 
\left({1\over \not\! k{\bf 1}_8
-{\bf M}_C P_L-{\bf M}_C^\dag P_R}\right)_{\tilde H_d^\delta \tilde W^k}P_L
\left({\bf h}_d\right)_{il}
\left(i\sigma_2\right)^{\delta\gamma}
\left({1\over k^2{\bf 1}_{12}-\widetilde{\bf M}^2}\right)_{\tilde Q_l^{\gamma}
\tilde Q_j^{\dag\beta}}\ \left(\frac{g_2\sigma_k^{\beta\alpha}}{\sqrt{2}}\right)
\nonumber\\
&&
\hspace{-2cm}
\left. + P_L
\left(\frac{1}{\not\! k{\bf 1}_8-{\bf M}_C P_L-{\bf M}_C^\dag P_R}
\right)_{\tilde B \tilde H_d^\beta}
\left(\frac{2 g_1}{3\sqrt{2}}\right)
P_L\left({1\over k^2{\bf 1}_{12}-\widetilde{\bf M}^2}\right)_{\tilde D_i
\tilde D_l^\dag}
\left({\bf h}_d\right)_{lj}
\left(i\sigma_2\right)^{\beta\alpha}
\right] \ ,\nonumber
\end{eqnarray}
where $\sigma_k$  are the usual  Pauli matrices, $P_{L(R)}=\frac{1}{2}
\left[1-(+)\gamma_5\right]$  is the left-handed  (right-handed) chiral
projection operator  and $g_{1,2,3}$ are the  gauge coupling constants
of $U(1)_Y,\ SU(2)_L$ and  $SU(3)_c$, respectively. Also, $C_F$ is the
quadratic Casimir invariant of  QCD in the fundamental representation,
i.e., $C_F=4/3$, and  ${\bf 1}_N$ is the $N\times  N$ identity matrix.
The $12\times 12$ squark mass-squared matrix $\widetilde{\bf M}^2$ and
the $8\times 8$ chargino-neutralino  mass matrix ${\bf M}_C$ are given
in     Appendix~\ref{app_masses}.       The     Greek     superscripts
in~(\ref{complete_down}) represent $SU(2)_L$ indices.

We  may  obtain   useful  approximations  to~(\ref{complete_down})  by
expanding in powers of the Higgs field~$\Phi_{1,2}$, so that
\begin{equation}
\label{down_expansion}
\mbox{\boldmath$\Delta$}{\bf h}_d\
\simeq\
{\bf h}_d\ \Big<\mbox{\boldmath$\Delta$}_d^{\Phi_1}
\Big>_0\ \Phi_1^\dag\ 
+\: {\bf h}_d\ \Big<\mbox{\boldmath$\Delta$}_d^{\Phi_2}
\Big>_0\ \Phi_2^\dag + \ldots\ ,
\end{equation}
where  $\big<   \ldots  \big>_0$  indicates  the   value  setting  all
background  fields to  zero in  the expression  enclosed, and  we have
introduced
\begin{equation}
\label{firstDelta}
\mbox{\boldmath$\Delta$}_d^{\Phi_i}\
\equiv\
{\bf h}_d^{-1}\;
\frac{\delta\mbox{\boldmath$\Delta$}{\bf h}_d}{\delta\Phi_i^\dag}\ .
\end{equation}
The higher-order terms correspond  to Feynman diagrams with additional
Higgs insertions  along the internal propagators,  which are typically
suppressed by additional factors  of $(M_{\rm EW}/M_{\rm SUSY})^2$ and
may safely be  neglected.  We call  the first term of such an expansion
the ``single-Higgs-insertion'' (SHI) approximation.

Working in  the SHI  approximation and assuming  flavour-diagonal soft
SUSY-breaking  terms, (\ref{complete_down}) may  be written  using the
expansion (\ref{down_expansion}) in the form
\begin{eqnarray}
\label{SE_approx_down_phi2}
\hspace{-1.0cm}
\Big<\mbox{\boldmath$\Delta$}_d^{\Phi_2}\Big>_0
& = &
{\bf 1}\ \frac{2\alpha_3}{3\pi}\ \mu^\ast M_3^\ast
\ {\it I}\left( \widetilde M_Q^2, \widetilde M_D^2, | M_3|^2 \right)
-
{\bf 1}\ \frac{\alpha_1}{36\pi}\ \mu^\ast M_1^\ast
\ {\it I}\left( \widetilde M_Q^2, \widetilde M_D^2, | M_1|^2 \right)
\nonumber\\
& & 
\hspace{-0.5cm}
+\ \ \ \frac{{\bf h}_u^\dag{\bf h}_u}{16\pi^2}\ \mu^\ast A_u^\ast
\ {\it I}\left( \widetilde M_Q^2, \widetilde M_U^2, |\mu|^2 \right)
\ \
-{\bf 1}\ \frac{3\alpha_2}{8\pi}\ \mu^\ast M_2^\ast
\ {\it I}\left( \widetilde M_Q^2, |M_2|^2, |\mu|^2 \right)\\
& &
\hspace{-0.5cm}
-\ {\bf 1}\ \frac{\alpha_1}{24\pi}\ \mu^\ast M_1^\ast
\ {\it I}\left( \widetilde M_Q^2, |M_1|^2, |\mu|^2 \right)
\ -\ 
{\bf 1}\ \frac{\alpha_1}{12\pi}\ \mu^\ast M_1^\ast
\ {\it I}\left( \widetilde M_D^2, |M_1|^2, |\mu|^2 \right),\ 
\nonumber\\
& &\nonumber\\
\label{SE_approx_down_phi1}
\hspace{-1.0cm}
\Big<\mbox{\boldmath$\Delta$}_d^{\Phi_1}\Big>_0
& = &
-{\bf 1}\ \frac{2\alpha_3}{3\pi}\ A_d M_3^\ast
\ {\it I}\left( \widetilde M_Q^2, \widetilde M_D^2, | M_3|^2 \right)
+
{\bf 1}\ \frac{\alpha_1}{36\pi}\ A_d M_1^\ast
\ {\it I}\left( \widetilde M_Q^2, \widetilde M_D^2, | M_1|^2 \right)
\nonumber\\
& & 
-\ \ \ \frac{{\bf h}_u^\dag{\bf h}_u}{16\pi^2}\ |\mu|^2
\ {\it I}\left( \widetilde M_Q^2, \widetilde M_U^2, |\mu|^2 \right)
\hspace{0.5cm}
+
{\bf 1}\ \frac{3\alpha_2}{8\pi}
\ {\it B}_0\left(0,|M_2|^2,M_Q^2 \right)\\
& &
+{\bf 1}\ \frac{\alpha_1}{24\pi}\ {\it B}_0\left(0,|M_1|^2,M_Q^2 \right)
\hspace{0.5cm}
+{\bf 1}\ \frac{\alpha_1}{12\pi}\ {\it B}_0\left(0,|M_1|^2,M_D^2 \right) ,
\nonumber
\end{eqnarray}
where ${\bf 1}\equiv{\bf 1}_3$ and $\alpha_i=g^2_i/4\pi$ as usual.  In
writing                                down~(\ref{SE_approx_down_phi2})
and~(\ref{SE_approx_down_phi1}), we have made  use of the the one-loop
function ${\it I}(a,b,c)$, given by
\begin{equation}
{\it I}(a,b,c)\ =\
\frac{ab\ln(a/b) + bc\ln(b/c) + ac\ln(c/a)}
{(a-b)(b-c)(a-c)}\ ,
\end{equation}
and the  Passarino-Veltman function  ${\it B}_0(p,a,b)$, which  may be
written as
\begin{equation}
B_0(0,a,b)\ =\ 1 - \ln\left(\frac{b}{Q^2}\right)
+\frac{a}{a-b}\ln\left(\frac{b}{a}\right)\ ,
\end{equation}
when  the first  argument $p^2$  is set  to zero,  i.e.~$p^2=0$.  Here
$Q^2$       is       the       renormalisation       scale.        The
approximation~(\ref{SE_approx_down_phi2})  agrees  with other  results
from   the  literature~\cite{Carena:1999py,Uli}  in   the  appropriate
limits.

In      both~(\ref{complete_down})       and      the      approximate
expressions~(\ref{SE_approx_down_phi2})
and~(\ref{SE_approx_down_phi1}), the  first two contributions  are the
gluino-   and  bino-mediated   corrections  of   the  form   shown  in
Fig.~\ref{CTE_down}(a), whereas the  third term represents the charged
Higgsino diagram  of Fig.~\ref{CTE_down}(b).   We note that  under the
assumption  of Minimal Flavour  Violation, this  Higgsino term  is the
only correction  with a non-trivial flavour structure  at the one-loop
level.    The  fourth   and  fifth   terms  give,   respectively,  the
contributions due  to wino-  and bino- exchange  diagrams of  the type
shown in Fig.~\ref{CTE_down}(c), whereas the  final term is due to the
bino-exchange  diagram of Fig.~\ref{CTE_down}(d).   We note  that this
final contribution  is the  only one that  is independent of  the soft
SUSY-breaking left-handed squark mass, $\widetilde M_Q^2$.

However, as discussed in  Section~\ref{sec:MFV}, the RG running of the
soft-SUSY  breaking  parameters  provides  an  additional  source  for
flavour  violation.    To  leading  order  in   the  shift  parameters
$\mbox{\boldmath$\delta$}    {\bf    \widetilde   M}^2_{Q,U,D}$    and
$\mbox{\boldmath$\delta$}{\bf              a}_{u,d}$             given
in~(\ref{MFVshift})~\footnote{A similar  expansion, albeit non-flavour
covariant,  was also  considered in~\cite{Durmus}.},  we find  that the
threshold       corrections       (\ref{SE_approx_down_phi2})      and
(\ref{SE_approx_down_phi1}) are modified by amounts
\begin{eqnarray}
  \label{ThresholdShift_down_phi2}
\Big<\mbox{\boldmath$\delta\Delta$}_d^{\Phi_2}\Big>_0
\!\!\!&=&\!\!
\frac{2\alpha_3}{3\pi}\; \mu^\ast M_3^\ast \left[\,
  \mbox{\boldmath$\delta$}{\bf \widetilde M}^2_Q\, 
{\it K}\left(\widetilde M_Q^2, \widetilde M_D^2, | M_3|^2 \right)\:
+\: {\bf h}^{-1}_d  \mbox{\boldmath$\delta$}{\bf \widetilde M}^2_D
{\bf h}_d\, {\it K}\left(\widetilde M_D^2, \widetilde M_Q^2, | M_3|^2
\right)\,\right] 
\nonumber\\
\!&&\!\hspace{-1.4cm}
-\ \frac{\alpha_1}{36\pi}\ \mu^\ast M_1^\ast \left[\,
  \mbox{\boldmath$\delta$}{\bf \widetilde M}^2_Q\, 
{\it K}\left(\widetilde M_Q^2, \widetilde M_D^2, | M_1|^2 \right)\:
+\: {\bf h}^{-1}_d  \mbox{\boldmath$\delta$}{\bf \widetilde M}^2_D
{\bf h}_d\, {\it K}\left(\widetilde M_D^2, \widetilde M_Q^2, | M_1|^2
\right)\,\right] 
\nonumber\\
\!&&\! \hspace{-1.4cm}
+\ \frac{1}{16\pi^2} \mu^\ast A_u^\ast
\left[\, {\bf h}_u^\dag \mbox{\boldmath$\delta$}{\bf \widetilde M}^2_U
{\bf h}_u \,
{\it K}\left(\widetilde M_U^2, \widetilde M_Q^2, |\mu|^2 \right)
+\: \mbox{\boldmath$\delta$}{\bf \widetilde M}^2_Q {\bf h}_u^\dag
{\bf h}_u\, {\it K}\left(\widetilde M_Q^2, \widetilde M_U^2, |\mu|^2
\right)\,\right]
\nonumber\\
\!&&\! \hspace{-1.4cm}
+\ \frac{\mbox{\boldmath$\delta$}{\bf a}_u^\dag\,{\bf
    h}_u}{16\pi^2}\ \mu^\ast\, 
{\it I}\left( \widetilde M_Q^2, \widetilde M_U^2, |\mu|^2 \right)\  -\ 
\frac{3\alpha_2}{8\pi}\ \mu^\ast M_2^\ast\, 
\mbox{\boldmath$\delta$}{\bf \widetilde M}^2_Q\ 
{\it K}\left( \widetilde M_Q^2, |M_2|^2, |\mu|^2 \right)\\
\!&&\!\hspace{-1.4cm}
-\ \frac{\alpha_1}{24\pi}\; \mu^\ast M_1^\ast\,
\mbox{\boldmath$\delta$}{\bf \widetilde M}^2_Q\, 
{\it K}\left( \widetilde M_Q^2, |M_1|^2, |\mu|^2 \right)\ -\ 
\frac{\alpha_1}{12\pi}\ \mu^\ast M_1^\ast\,
{\bf h}^{-1}_d  \mbox{\boldmath$\delta$}{\bf \widetilde M}^2_D
{\bf h}_d\, {\it K}\left( \widetilde M_D^2, |M_1|^2, |\mu|^2 \right)\;,
\nonumber\\
& &\nonumber\\
  \label{ThresholdShift_down_phi1}
\Big<\mbox{\boldmath$\delta\Delta$}_d^{\Phi_1}\Big>_0
\!\!\!&=&\!\!
-\frac{2\alpha_3}{3\pi}\; A_d M_3^\ast\, \left[\,
  \mbox{\boldmath$\delta$}{\bf \widetilde M}^2_Q\, 
{\it K}\left(\widetilde M_Q^2, \widetilde M_D^2, | M_3|^2 \right)\:
+\: {\bf h}^{-1}_d  \mbox{\boldmath$\delta$}{\bf \widetilde M}^2_D
{\bf h}_d\, {\it K}\left(\widetilde M_D^2, \widetilde M_Q^2, | M_3|^2
\right)\,\right] 
\nonumber\\
\!&&\!\hspace{-1.4cm}
-\frac{2\alpha_3}{3\pi}\;{\bf h}_d^{-1}\mbox{\boldmath$\delta$}{\bf
  a}_d M_3^\ast\, 
{\it I}\left( \widetilde M_Q^2, \widetilde M_D^2, |M_3|^2 \right)
+\frac{\alpha_1}{36\pi}\;{\bf h}_d^{-1}\mbox{\boldmath$\delta$}{\bf
  a}_d M_1^\ast\, 
{\it I}\left( \widetilde M_Q^2, \widetilde M_D^2, |M_1|^2 \right)
\nonumber\\
\!&&\!\hspace{-1.4cm}
+\frac{\alpha_1}{36\pi}\; A_d M_1^\ast\, \left[\,
  \mbox{\boldmath$\delta$}{\bf \widetilde M}^2_Q\, 
{\it K}\left(\widetilde M_Q^2, \widetilde M_D^2, | M_1|^2 \right)\:
+\: {\bf h}^{-1}_d  \mbox{\boldmath$\delta$}{\bf \widetilde M}^2_D
{\bf h}_d\, {\it K}\left(\widetilde M_D^2, \widetilde M_Q^2, |M_1|^2
\right)\,\right] 
\nonumber\\
\!&&\! \hspace{-1.4cm}
-\ \frac{1}{16\pi^2} |\mu|^2 
\left[\, {\bf h}_u^\dag \mbox{\boldmath$\delta$}{\bf \widetilde M}^2_U
{\bf h}_u \,
{\it K}\left(\widetilde M_U^2, \widetilde M_Q^2, |\mu|^2 \right)
+\: \mbox{\boldmath$\delta$}{\bf \widetilde M}^2_Q {\bf h}_u^\dag
{\bf h}_u\, {\it K}\left(\widetilde M_Q^2, \widetilde M_U^2, |\mu|^2
\right)\,\right]
\\
\!&&\! \hspace{-1.4cm}
+\frac{3\alpha_2}{8\pi}\mbox{\boldmath$\delta$}{\bf \widetilde M}^2_Q\,
{\it I}\left( \widetilde M_Q^2, |M_2|^2 \right)
+\frac{\alpha_1}{24\pi}\left[
\mbox{\boldmath$\delta$}{\bf \widetilde M}^2_Q\,
{\it I}\left( \widetilde M_Q^2, |M_1|^2 \right)
+2{\bf h}_d^{-1}
\mbox{\boldmath$\delta$}{\bf \widetilde M}^2_D{\bf h}_d\,
{\it I}\left( \widetilde M_D^2, |M_1|^2 \right)\right]\ ,
\nonumber
\end{eqnarray}
where
\begin{eqnarray}
  \label{Kabc}
{\it K}(a,b,c)\
& = &
\ \frac{d}{da}\; {\it I}(a,b,c)
\nonumber\\
& = &\ 
\frac{b\ln\left(a/b\right)+c\ln\left(c/a\right)}
{\left(a-b\right)\left(b-c\right)\left(a-c\right)}\
+\ \frac{\left(b+c-2a\right){\it I}\left( a, b, c \right)+1}
{\left(a-b\right)\left(a-c\right)}\; ,
\end{eqnarray}
and
\begin{eqnarray}
  \label{Iab}
{\it I}(a,b) & \equiv &
\frac{d}{da}\; {\it B}_0(0,a,b)\ =\
-\,\lim_{c\to a} {\it I}(a,b,c)\nonumber\\
& = &
\frac{1}{a-b}\; \left[\,\frac{b}{a-b}\ln\left(\frac{a}{b}\right)-1\,\right]\ .
\end{eqnarray}

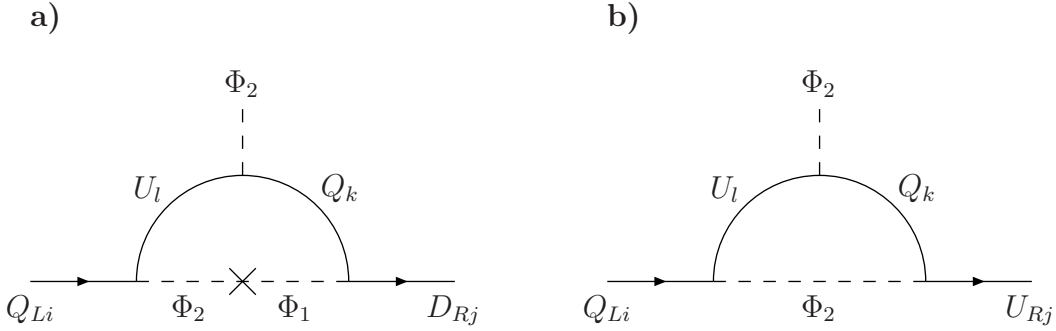
\begin{figure}[t!]
\begin{center}
\begin{tabular}{ccc}
\begin{picture}(180,150)(0,0)
\ArrowLine(10,20)(50,20)
\DashLine(50,20)(130,20){5}
\ArrowLine(130,20)(170,20)
\CArc(90,20)(40,0,180)
\DashLine(90,60)(90,85){5}
\Text(90,90)[b]{$\Phi_2$}
\Line(85,15)(95,25)
\Line(95,15)(85,25)
\Text(10,15)[t]{$Q_{Li}$}
\Text(170,15)[t]{$D_{Rj}$}
\Text(60,50)[rb]{$U_l$}
\Text(120,50)[lb]{$Q_k$}
\Text(70,15)[t]{$\Phi_2$}
\Text(110,15)[t]{$\Phi_1$}
\Text(10,120)[l]{\bf a)}
\end{picture}
& \hspace{0.5 cm} &
\begin{picture}(180,150)(0,0)
\ArrowLine(10,20)(50,20)
\DashLine(50,20)(130,20){5}
\ArrowLine(130,20)(170,20)
\CArc(90,20)(40,0,180)
\DashLine(90,60)(90,85){5}
\Text(90,90)[b]{$\Phi_2$}
\Text(10,15)[t]{$Q_{Li}$}
\Text(170,15)[t]{$U_{Rj}$}
\Text(60,50)[rb]{$U_l$}
\Text(120,50)[lb]{$Q_k$}
\Text(90,15)[t]{$\Phi_2$}
\Text(10,120)[l]{\bf b)}
\end{picture}
\end{tabular}
\caption{\it 2HDM contribution to the down-type (a) and up-type (b) quark
self-energies, to first order in $\Phi_2$.}
\label{2HDM}
\end{center}
\end{figure}

\noindent
There  is also a  two-Higgs-doublet model  (2HDM) contribution  to the
one-loop  self-energy graphs  for down-type  quarks which  is formally
$\tan\beta$-enhanced.      This     contribution,     displayed     in
Fig.~\ref{2HDM}(a),  is  not  affected  by flavour non-universal RG  effects  of  the  sort
discussed above.  It may be calculated by evaluating
\begin{eqnarray}
\label{complete_2hdm_down}
\left(\mbox{\boldmath$\Delta$}{\bf h}_d^{\rm 2HDM}\right)_{ij} & = &
\int \frac{d^n k}{(2\pi)^n i}
\left(\bf{h}_d\right)_{il} P_L
\left(\frac{1}{\not\! k{\bf 1}_6 - {\bf M}_q P_L - {\bf M}_q^\dag P_R}\right)
_{Q_l u_k} P_L \left({\bf h}_u\right)_{kj}
\nonumber\\
& &
\times \left(\frac{1}{k^2{\bf 1}_4 - {\bf M}_H^2}\right)_{\Phi_1 \Phi_2^\dag}
\, ,
\end{eqnarray}
where  ${\bf M}_q$  is the  $6\times 6$  quark mass  matrix  and ${\bf
M}_H^2$ the $8\times 8$ Higgs mass-squared matrix.  Explicit forms for
these  matrices  are   given  in  Appendix~\ref{app_masses}.   In  the
SHI   approximation,  the   2HDM   contribution  of
(\ref{complete_2hdm_down}) may be written as
\begin{equation}
(\mbox{\boldmath$\Delta$}_d^{\Phi_2})^{\rm 2HDM}\
=\
\frac{{\bf h}_u^\dag{\bf h}_u}{16\pi^2}
\frac{B^\ast \mu^\ast}{M^2_{H_d}- M^2_{H_u}}
\ln\left|\frac{M^2_{H_d}+|\mu|^2}{M^2_{H_u}+|\mu|^2}\right|
\ .
\end{equation}
The corresponding term $(\mbox{\boldmath$\Delta$}_d^{\Phi_1})^{\rm 2HDM}$ 
for $\Phi_1$ is zero at this level of approximation.

\subsubsection{Up-type Quark Yukawa Couplings}

\begin{figure}[t!]
\begin{center}
\begin{tabular}{ccc}
\begin{picture}(180,150)(0,0)
\ArrowLine(10,50)(50,50)
\Line(50,50)(130,50)
\ArrowLine(130,50)(170,50)
\Photon(50,50)(130,50){4}{7}
\DashCArc(90,50)(40,0,180){5}
\DashLine(90,90)(90,115){5}
\Text(90,120)[b]{$\Phi_2$}
\Line(85,45)(95,55)
\Line(95,45)(85,55)
\Text(10,45)[t]{$Q_{Li}$}
\Text(170,45)[t]{$U_{Rj}$}
\Text(60,80)[rb]{$\tilde Q_i$}
\Text(120,80)[lb]{$\tilde U_j$}
\Text(90,40)[t]{$\tilde g\ ,\ \widetilde B$}
\Text(10,120)[l]{\bf a)}
\end{picture}
& \hspace{0.5 cm} &
\begin{picture}(180,150)(0,0)
\ArrowLine(10,50)(50,50)
\Line(50,50)(130,50)
\ArrowLine(130,50)(170,50)
\DashCArc(90,50)(40,0,180){5}
\DashLine(90,90)(90,115){5}
\Text(90,120)[b]{$\Phi_2$}
\Line(85,45)(95,55)
\Line(95,45)(85,55)
\Text(10,45)[t]{$Q_{Li}$}
\Text(170,45)[t]{$U_{Rj}$}
\Text(60,80)[rb]{$\tilde D_l$}
\Text(120,80)[lb]{$\tilde Q_k$}
\Text(70,45)[t]{$\tilde h_d$}
\Text(110,45)[t]{$\tilde h_u$}
\Text(10,120)[l]{\bf b)}
\end{picture}
\\
\begin{picture}(180,150)(0,0)
\ArrowLine(10,50)(50,50)
\Line(50,50)(130,50)
\ArrowLine(130,50)(170,50)
\Photon(50,50)(90,50){4}{4}
\DashCArc(90,50)(40,0,180){5}
\DashLine(90,50)(90,20){5}
\Text(90,15)[t]{$\Phi_2$}
\Line(65,45)(75,55)
\Line(75,45)(65,55)
\Line(85,45)(95,55)
\Line(95,45)(85,55)
\Text(90,95)[b]{$\tilde Q_{i}$}
\Text(70,42)[t]{$\widetilde W^k,\widetilde B$}
\Text(110,45)[t]{$\tilde h_u$}
\Text(10,45)[t]{$Q_{Li}$}
\Text(170,45)[t]{$U_{Rj}$}
\Text(10,120)[l]{\bf c)}
\end{picture}
& \hspace{0.5 cm} &
\begin{picture}(180,150)(0,0)
\ArrowLine(10,50)(50,50)
\Line(50,50)(130,50)
\ArrowLine(130,50)(170,50)
\Photon(90,50)(130,50){4}{4}
\DashCArc(90,50)(40,0,180){5}
\DashLine(90,50)(90,20){5}
\Text(90,15)[t]{$\Phi_2$}
\Line(85,45)(95,55)
\Line(95,45)(85,55)
\Line(105,45)(115,55)
\Line(115,45)(105,55)
\Text(90,95)[b]{$\tilde U_{j}$}
\Text(110,42)[t]{$\widetilde B$}
\Text(70,45)[t]{$\tilde h_u$}
\Text(10,45)[t]{$Q_{Li}$}
\Text(170,45)[t]{$U_{Rj}$}
\Text(10,120)[l]{\bf d)}
\end{picture}
\end{tabular}
\caption{\it The complete set of gauge- and flavour-covariant diagrams
contributing to the up-type quark self-energy, to first order in
$\Phi_2$.  Note that in panels (a) and (c) contributions from all listed
gauginos should be included.}
\label{CTE_up}
\end{center}
\end{figure}
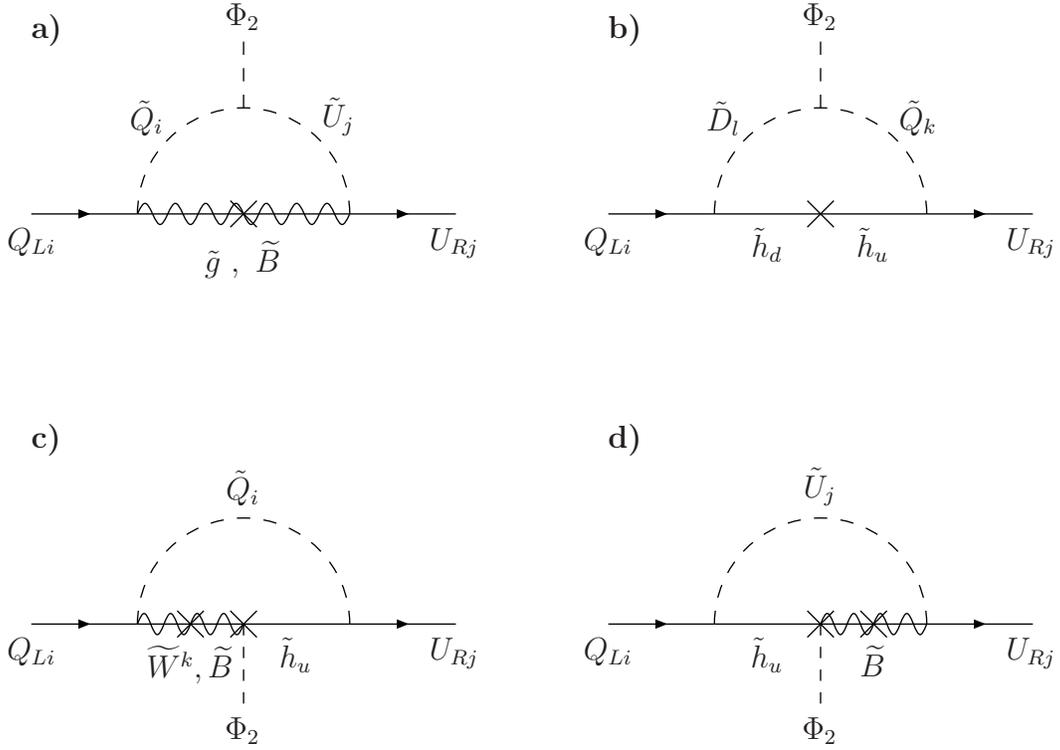

We now turn  our attention to the up-quark  sector.  The up-type quark
self-energy is described by the effective Lagrangian
\begin{equation}
\label{L_eff_up}
-{\mathcal L}^u_{\rm eff}\left[\Phi_1,\Phi_2\right]\
=\
\bar{u}^0_{iR}
\left({\bf h}_u \Phi_2^{T\alpha}
+\mbox{\boldmath$\Delta$}{\bf h}_u^\alpha\left[\Phi_1,\Phi_2\right]\right)_{ij}
\left(-i\sigma_2\right)^{\alpha\beta}
Q^{0\beta}_{jL}\ ,
\end{equation}
where
\begin{equation}
\mbox{\boldmath$\Delta$}{\bf h}_u = \mbox{\boldmath$\Delta$}{\bf h}_u^{\rm 2HDM}
+ \mbox{\boldmath$\Delta$}{\bf h}_u^{\rm SUSY}
+ \mbox{\boldmath$\Delta$}{\bf h}_u^{\rm CT}\ ,
\end{equation}
and  $\mbox{\boldmath$\Delta$}{\bf  h}_u^{\rm   CT}$  is  due  to  the
wave-function renormalization of the Higgs and quark superfields.  The
corrections to the  up-type quark self-energy are given  by the set of
diagrams  displayed  in   Fig.~\ref{CTE_up}.   We  may  express  these
corrections as
\begin{eqnarray}
\label{complete_up}
-\left(\mbox{\boldmath$\Delta$}{\bf h}_u^{\rm SUSY}\right)^\alpha_{ij}
& = & 
\int {d^n k\over(2\pi)^n i}
\left[P_L {-2C_F\,g_3^2 M_3^\ast\over k^2-|M_3^2|}
\left({1\over k^2{\bf 1}_{12}-\widetilde{\bf M}^2}\right)_
{\tilde U_i\tilde Q_j^{\dag\beta}}
\left(i\sigma_2\right)^{\beta\alpha}
\right.\nonumber\\
&&
\hspace{-3cm}
+\ P_L \left(\frac{-2g_1^2}{9}\right)
\left({1\over \not\! k {\bf 1}_8-{\bf M}_C P_L
-{\bf M}_C^\dag P_R}\right)_{\tilde B \tilde B}
P_L
\left({1\over k^2{\bf 1}_{12}-\widetilde{\bf M}^2}\right)_
{\tilde U_i\tilde Q_j^{\dag\beta}}
\left(i\sigma_2\right)^{\beta\alpha}
\nonumber\\
&&
\hspace{-3cm}
+\ P_L\left({1\over \not\! k {\bf 1}_8-{\bf M}_C P_L
-{\bf M}_C^\dag P_R}\right)_{\tilde H_u^\gamma \tilde H_d^\alpha}
P_L \left({\bf h}_u\right)_{il}
\left(i\sigma_2\right)^{\gamma\beta}
\left({1\over k^2{\bf 1}_{12}-\widetilde{\bf M}^2}\right)_
{\tilde Q_l^\beta \tilde D_k^\dag} \left({\bf h}_d\right)_{kj}\\
&&
\hspace{-3cm}
+ P_L 
\left({1\over \not\! k{\bf 1}_8
-{\bf M}_C P_L-{\bf M}_C^\dag P_R}\right)_{\tilde H_u^\delta \tilde B}P_L
\left({\bf h}_u\right)_{il}
\left(-i\sigma_2\right)^{\delta\gamma}
\left({1\over k^2{\bf 1}_{12}-\widetilde{\bf M}^2}\right)_{\tilde Q_l^\gamma
\tilde Q_j^{\dag\beta}}\left(\frac{g_1}{3\sqrt{2}}\right)
\left(i\sigma\right)^{\beta\alpha}
\nonumber\\
&&
\hspace{-3cm}
+\sum_k P_L 
\left({1\over \not\! k{\bf 1}_8
-{\bf M}_C P_L-{\bf M}_C^\dag P_R}\right)_{\tilde H_u^\epsilon \tilde W^k}P_L
\left({\bf h}_u\right)_{il}
\left(-i\sigma_2\right)^{\epsilon\delta}
\left({1\over k^2{\bf 1}_{12}-\widetilde{\bf M}^2}\right)_{\tilde Q_l^\delta
\tilde Q_j^{\dag\gamma}}\ 
\left(\frac{g_2\sigma_k^{\gamma\beta}}{\sqrt{2}}\right)
\left(i\sigma\right)^{\beta\alpha}
\nonumber\\
&&
\hspace{-3cm}
\left. + P_L
\left(\frac{1}{\not\! k{\bf 1}_8-{\bf M}_C P_L-{\bf M}_C^\dag P_R}
\right)_{\tilde B \tilde H_u^\alpha}
\left(\frac{-4g_1}{3\sqrt{2}}\right)
P_L\left({1\over k^2{\bf 1}_{12}-\widetilde{\bf M}^2}\right)_{\tilde U_i
\tilde U_l^\dag}
\left({\bf h}_u\right)_{lj}
\right]\ .\nonumber
\end{eqnarray}

\noindent
Expanding~(\ref{complete_up}) in powers of $\Phi_{1,2}$, we may write
\begin{equation}
\label{up_expansion}
\mbox{\boldmath$\Delta$}{\bf h}_u\
\simeq\
{\bf h}_u\ \Big<\mbox{\boldmath$\Delta$}_u^{\Phi_1}\Big>_0
\Phi_1^T\: +\:
{\bf h}_u\ \Big<\mbox{\boldmath$\Delta$}_u^{\Phi_2}\Big>_0 \Phi_2^T\:
+\: \ldots\ ,
\end{equation}
with
\begin{equation}
\mbox{\boldmath$\Delta$}_u^{\Phi_i}\
=\
{\bf h}_u^{-1}\:
\frac{\delta\mbox{\boldmath$\Delta$}{\bf h}_u}{\delta\Phi_i^T}\ ,
\end{equation}
so that the up-type quark self-energy is given in the SHI approximation by
\begin{eqnarray}
\label{SE_approx_up_phi2}
\hspace{-1.0cm}
\Big<\mbox{\boldmath$\Delta$}_u^{\Phi_2}\Big>_0
& = &
-{\bf 1}\ \frac{2\alpha_3}{3\pi}\ A_u M_3^\ast
\ {\it I}\left( \widetilde M_Q^2, \widetilde M_U^2, | M_3|^2 \right)
-
{\bf 1}\ \frac{\alpha_1}{18\pi}\ A_u M_1^\ast
\ {\it I}\left( \widetilde M_Q^2, \widetilde M_U^2, | M_1|^2 \right)
\nonumber\\
& &
-\ \ \ \frac{{\bf h}_d^\dag{\bf h}_d}{16\pi^2}\ |\mu|^2
\ {\it I}\left( \widetilde M_Q^2, \widetilde M_D^2, |\mu|^2 \right)
\ \ 
+{\bf 1}\ \frac{3\alpha_2}{8\pi}\ 
\ {\it B}_0\left(0, \widetilde M_Q^2, |M_2|^2 \right)\\
& &
\ -{\bf 1}\ \frac{\alpha_1}{24\pi}
\ {\it B}_0\left(0, |M_1|^2, \widetilde M_Q^2 \right)
\ +\ 
{\bf 1}\ \frac{\alpha_1}{6\pi}\ 
\ {\it B}_0\left(0, |M_1|^2, \widetilde M_U^2 \right),\ 
\nonumber\\
& &\nonumber\\
\label{SE_approx_up_phi1}
\hspace{-1.0cm}
\Big<\mbox{\boldmath$\Delta$}_u^{\Phi_1}\Big>_0
& = &
{\bf 1}\ \frac{2\alpha_3}{3\pi}\ \mu^\ast M_3^\ast
\ {\it I}\left(M_Q^2,M_U^2,|M_3|^2\right)
+
{\bf 1}\ \frac{\alpha_1}{18\pi}\ \mu^\ast M_1^\ast 
\ {\it I}\left(M_Q^2, M_U^2, |M_1|^2\right)\nonumber\\
& &
+\ \ \ \frac{{\bf h}_d^\dag{\bf h}_d}{16\pi^2}\ \mu^\ast A_d^\ast
\ {\it I}\left(M_Q^2, M_D^2, |\mu|^2\right)
-\ \frac{3\alpha_2}{8\pi}\ \mu^\ast M_2^\ast
\ {\it I}\left(M_Q^2, |M_2|^2, |\mu|^2 \right)\\
& &
+{\bf 1}\ \frac{\alpha_1}{24\pi}\ \mu^\ast M_1^\ast
\ {\it I}\left(M_Q^2, |M_1|^2, |\mu|^2\right)
-{\bf 1}\ \frac{\alpha_1}{6\pi}\ \mu^\ast M_1^\ast
\ {\it I}\left(M_U^2, |M_1|^2, |\mu|^2\right)\ .
\nonumber
\end{eqnarray}
The RG running of the  soft-SUSY breaking parameters again provides an
additional source of flavour violation.  To leading order in the shift
parameters $\mbox{\boldmath$\delta$} {\bf \widetilde M}^2_{Q,U,D}$ and
$\mbox{\boldmath$\delta$}{\bf a}_{u,d}$  given in~(\ref{MFVshift}), we
find   that    the   threshold   corrections~(\ref{SE_approx_up_phi2})
and~(\ref{SE_approx_up_phi1}) are modified by
\begin{eqnarray}
\label{ThresholdShift_up_phi2}
\Big<\mbox{\boldmath$\delta$}{\bf\Delta}^{\Phi_2}_u\Big>_0
\!\!\!&=&\!\!
- \frac{2\alpha_3}{3\pi}\ A_u M_3^\ast\ \left[
\mbox{\boldmath$\delta$}\widetilde{\bf M}^2_Q
\ {\it K}\left( \widetilde M_Q^2, \widetilde M_U^2, | M_3|^2 \right)
+{\bf h}_u^{-1}\mbox{\boldmath$\delta$}\widetilde{\bf M}^2_U{\bf h}_u
\ {\it K}\left( \widetilde M_U^2, \widetilde M_Q^2, | M_3|^2 \right)
\right]
\nonumber\\
\!&&\!\hspace{-1.4cm}
 -\frac{2\alpha_3}{3\pi}\ {\bf h}_u^{-1}\mbox{\boldmath$\delta$}{\bf
   a}_u M_3^\ast\  
\ {\it I}\left( \widetilde M_Q^2, \widetilde M_U^2, | M_3|^2 \right)
-
\frac{\alpha_1}{18\pi}\ {\bf h}_u^{-1}\mbox{\boldmath$\delta$}{\bf a}_u M_1^\ast\ 
\ {\it I}\left( \widetilde M_Q^2, \widetilde M_U^2, | M_1|^2 \right)
\nonumber\\
\!&&\!\hspace{-1.4cm}
 -\frac{\alpha_1}{18\pi}\ A_u M_1^\ast\ \left[
\mbox{\boldmath$\delta$}\widetilde{\bf M}^2_Q
\ {\it K}\left( \widetilde M_Q^2, \widetilde M_U^2, | M_1|^2 \right)
+{\bf h}_u^{-1}\mbox{\boldmath$\delta$}\widetilde{\bf M}^2_U{\bf h}_u
\ {\it K}\left( \widetilde M_U^2, \widetilde M_Q^2, | M_1|^2 \right)
\right] 
\nonumber\\
\!&&\!\hspace{-1.4cm}
 -\frac{1}{16\pi^2} |\mu|^2 \left[
{\bf h}_d^\dag \mbox{\boldmath$\delta$}\widetilde{\bf M}^2_D {\bf h}_d
\ {\it K}\left( \widetilde M_D^2, \widetilde M_Q^2, |\mu|^2 \right)
+\mbox{\boldmath$\delta$}\widetilde{\bf M}^2_Q {\bf h}_d^\dag{\bf h}_d
\ {\it K}\left( \widetilde M_Q^2, \widetilde M_D^2, |\mu|^2 \right)
\right] 
\\
\!&&\!\hspace{-1.4cm}
 +\frac{3\alpha_2}{8\pi}\mbox{\boldmath$\delta$}\widetilde{\bf M}^2_Q
\ {\it I}\left(\widetilde M_Q^2, |M_2|^2\right)
-\frac{\alpha_1}{24\pi}\left[\mbox{\boldmath$\delta$}\widetilde{\bf M}^2_Q
\ {\it I}\left(\widetilde M_Q^2, |M_1|^2\right)
-4{\bf h}_u^{-1}\mbox{\boldmath$\delta$}\widetilde{\bf M}^2_U{\bf h}_u
\ {\it I}\left(\widetilde M_U^2, |M_1|^2\right)\right]\ .
\nonumber\\
& &\nonumber\\
\label{ThresholdShift_up_phi1}
\Big<\mbox{\boldmath$\delta$}{\bf\Delta}^{\Phi_1}_u\Big>_0
\!\!\!&=&\!\!
\frac{2\alpha_3}{3\pi}\ \mu^\ast M_3^\ast\ \left[
\mbox{\boldmath$\delta$}\widetilde{\bf M}^2_Q
\ {\it K}\left( \widetilde M_Q^2, \widetilde M_U^2, | M_3|^2 \right)
+{\bf h}_u^{-1}\mbox{\boldmath$\delta$}\widetilde{\bf M}^2_U{\bf h}_u
\ {\it K}\left( \widetilde M_U^2, \widetilde M_Q^2, | M_3|^2 \right)
\right]
\nonumber\\
\!&&\!\hspace{-1.4cm}
+\frac{\alpha_1}{18\pi}\ \mu^\ast M_1^\ast\ \left[
\mbox{\boldmath$\delta$}\widetilde{\bf M}^2_Q
\ {\it K}\left( \widetilde M_Q^2, \widetilde M_U^2, | M_1|^2 \right)
+{\bf h}_u^{-1}\mbox{\boldmath$\delta$}\widetilde{\bf M}^2_U{\bf h}_u
\ {\it K}\left( \widetilde M_U^2, \widetilde M_Q^2, | M_1|^2 \right)
\right]
\nonumber\\
\!&&\!\hspace{-1.4cm}
+\frac{1}{16\pi^2} \ \mu^\ast A_d^\ast\ \left[
{\bf h}_d^\dag \mbox{\boldmath$\delta$}\widetilde{\bf M}^2_D {\bf h}_d
\ {\it K}\left( \widetilde M_D^2, \widetilde M_Q^2, |\mu|^2 \right)
+\mbox{\boldmath$\delta$}\widetilde{\bf M}^2_Q {\bf h}_d^\dag{\bf h}_d
\ {\it K}\left( \widetilde M_Q^2, \widetilde M_D^2, |\mu|^2 \right)
\right]
\nonumber\\
\!&&\!\hspace{-1.4cm}
+\frac{\mbox{\boldmath$\delta$}{\bf a}_d^\dag {\bf h}_d}{16\pi^2}
\mu^\ast\ {\it I}\left( \widetilde M_Q^2, \widetilde M_D^2, |\mu|^2 \right)
-
\frac{3\alpha_2}{8\pi}\ \mu^\ast M_2^\ast
\mbox{\boldmath$\delta$}\widetilde{\bf M}^2_Q 
\ {\it K}\left( \widetilde M_Q^2, |M_2|^2, |\mu|^2 \right)
\\
\!&&\!\hspace{-1.4cm}
+\frac{\alpha_1}{24\pi}\ \mu^\ast M_1^\ast
\mbox{\boldmath$\delta$}\widetilde{\bf M}^2_Q 
\ {\it K}\left( \widetilde M_Q^2, |M_1|^2, |\mu|^2 \right)
-\frac{\alpha_1}{6\pi}\ \mu^\ast M_1^\ast 
{\bf h}_u^{-1}\mbox{\boldmath$\delta$}\widetilde{\bf M}^2_U {\bf h}_u
\ {\it K}\left( \widetilde M_U^2, |M_1|^2, |\mu|^2 \right)\ .\nonumber
\end{eqnarray}
For completeness, we  also display the 2HDM correction  to the up-type
quark   self-energy  analogous   to~(\ref{complete_2hdm_down}).   This
contribution, as shown in Fig.~\ref{2HDM}(b), is a Standard Model-like
correction which  again is  not affected by  the RG  effects discussed
above.  It is given by
\begin{eqnarray}
\label{complete_2hdm_up}
\left(\mbox{\boldmath$\Delta$}{\bf h}_u^{\rm 2HDM}\right)_{ij} & = &
\int \frac{d^n k}{(2\pi)^n i}
\left(\bf{h}_u\right)_{il} P_L
\left(\frac{1}{\not\!{k}{\bf 1}_6 - {\bf M}_q P_L - {\bf M}_q^\dag P_R}\right)
_{Q_l u_k} P_L \left({\bf h}_u\right)_{kj}
\nonumber\\
& &
\times \left(\frac{1}{k^2{\bf 1}_4 - {\bf M}_H^2}\right)_{\Phi_2 \Phi_2^\dag}
\, .
\end{eqnarray}

\subsubsection{Lepton Yukawa Couplings}

\begin{figure}[t!]
\begin{center}
\begin{tabular}{ccc}
\begin{picture}(180,150)(0,0)
\ArrowLine(10,50)(50,50)
\Line(50,50)(130,50)
\ArrowLine(130,50)(170,50)
\Photon(50,50)(130,50){4}{7}
\DashCArc(90,50)(40,0,180){5}
\DashLine(90,90)(90,115){5}
\Text(90,120)[b]{$\Phi_2$}
\Line(85,45)(95,55)
\Line(95,45)(85,55)
\Text(10,45)[t]{$L_{Li}$}
\Text(170,45)[t]{$E_{Rj}$}
\Text(60,80)[rb]{$\tilde L_i$}
\Text(120,80)[lb]{$\tilde E_j$}
\Text(90,40)[t]{$\widetilde B$}
\Text(10,120)[l]{\bf a)}
\end{picture}
& \hspace{0.5 cm} &
\begin{picture}(180,150)(0,0)
\ArrowLine(10,50)(50,50)
\Line(50,50)(130,50)
\ArrowLine(130,50)(170,50)
\Photon(50,50)(90,50){4}{4}
\DashCArc(90,50)(40,0,180){5}
\DashLine(90,50)(90,20){5}
\Text(90,15)[t]{$\Phi_2$}
\Line(65,45)(75,55)
\Line(75,45)(65,55)
\Line(85,45)(95,55)
\Line(95,45)(85,55)
\Line(105,45)(115,55)
\Line(115,45)(105,55)
\Text(90,95)[b]{$\tilde L_{i}$}
\Text(70,42)[t]{$\widetilde B$}
\Text(102,45)[t]{$\tilde h_u$}
\Text(122,45)[t]{$\tilde h_d$}
\Text(10,45)[t]{$L_{Li}$}
\Text(170,45)[t]{$E_{Rj}$}
\Text(10,120)[l]{\bf b)}
\end{picture}
\\
\begin{picture}(180,150)(0,0)
\ArrowLine(10,50)(50,50)
\Line(50,50)(130,50)
\ArrowLine(130,50)(170,50)
\Photon(50,50)(90,50){4}{4}
\DashCArc(90,50)(40,0,180){5}
\DashLine(90,50)(90,20){5}
\Text(90,15)[t]{$\Phi_2$}
\Line(65,45)(75,55)
\Line(75,45)(65,55)
\Line(85,45)(95,55)
\Line(95,45)(85,55)
\Line(105,45)(115,55)
\Line(115,45)(105,55)
\Text(90,95)[b]{$\tilde L_{i}$}
\Text(70,42)[t]{$\widetilde W^k$}
\Text(102,45)[t]{$\tilde h_u$}
\Text(122,45)[t]{$\tilde h_d$}
\Text(10,45)[t]{$L_{Li}$}
\Text(170,45)[t]{$E_{Rj}$}
\Text(10,120)[l]{\bf c)}
\end{picture}
& \hspace{0.5 cm} &
\begin{picture}(180,150)(0,0)
\ArrowLine(10,50)(50,50)
\Line(50,50)(130,50)
\ArrowLine(130,50)(170,50)
\Photon(90,50)(130,50){4}{4}
\DashCArc(90,50)(40,0,180){5}
\DashLine(90,50)(90,20){5}
\Text(90,15)[t]{$\Phi_2$}
\Line(65,45)(75,55)
\Line(75,45)(65,55)
\Line(85,45)(95,55)
\Line(95,45)(85,55)
\Line(105,45)(115,55)
\Line(115,45)(105,55)
\Text(90,95)[b]{$\tilde E_{j}$}
\Text(110,42)[t]{$\widetilde B$}
\Text(62,45)[t]{$\tilde h_d$}
\Text(82,45)[t]{$\tilde h_u$}
\Text(10,45)[t]{$L_{Li}$}
\Text(170,45)[t]{$E_{Rj}$}
\Text(10,120)[l]{\bf d)}
\end{picture}
\end{tabular}
\caption{\it The complete set of gauge- and flavour-covariant diagrams
contributing to the lepton self-energy, to first order in
$\Phi_2$.}
\label{CTE_lepton}
\end{center}
\end{figure}
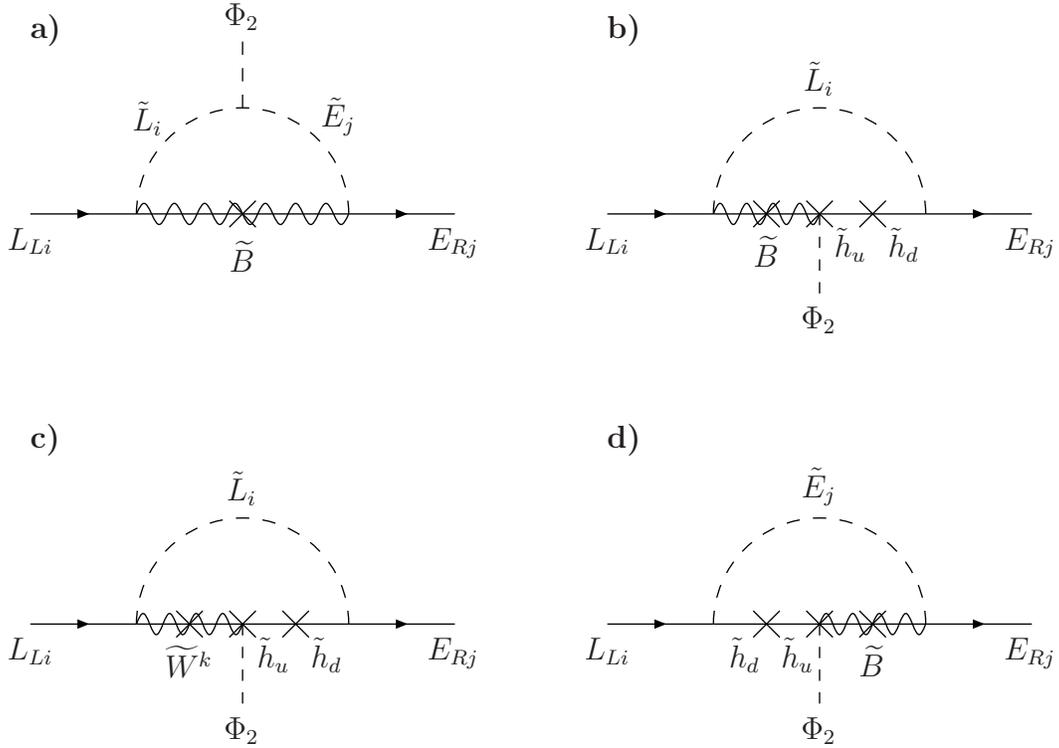

Finally, we  discuss the Higgs  couplings to the leptonic  sector.  In
analogy to~(\ref{L_eff_down}), we  may describe the lepton self-energy
by the effective Lagrangian
\begin{equation}
\label{L_eff_lepton}
-{\mathcal L}^e_{\rm eff}\left[\Phi_1,\Phi_2\right]\
=\
\bar{e}^0_{iR}
\left({\bf h}_e \Phi_1^{\dag\alpha}
+\mbox{\boldmath$\Delta$}{\bf h}_e^\alpha\left[\Phi_1,\Phi_2\right]\right)_{ij}
L^{0\alpha}_{jL}
+ {\rm H.c.}\ ,
\end{equation}
where   $\mbox{\boldmath$\Delta$}{\bf   h}_e$   is   a   $3\times   3$
Coleman-Weinberg effective  functional of the  background Higgs fields
which  encodes  the higher-order  corrections.   Only the  electroweak
corrections  contribute  to the  lepton  self-energy  at the  one-loop
level,    corresponding   to   the    Feynman   diagrams    shown   in
Fig.~\ref{CTE_lepton}.  Again separating out the one-loop counterterms
and neglecting  the 2HDM-like contributions, we  may write $\Delta{\bf
  h}_e^{\rm SUSY}$ as
\begin{eqnarray}
\label{complete_lepton}
-\left(\mbox{\boldmath$\Delta$}{\bf h}_e^{\rm SUSY}\right)_{ij}^\alpha & = & 
\int \frac{d^n k}{(2\pi)^n i}\ \left[
-\ P_L g_1^2 \left({1\over \not\! k {\bf 1}_8-{\bf M}_C P_L
-{\bf M}_C^\dag P_R}\right)_{\tilde B \tilde B}
P_L
\left({1\over k^2{\bf 1}_{12}-\widetilde{\bf M}_l^2}\right)_
{\tilde E_i\tilde L_j^{\dag\alpha}}\right. \nonumber\\
&&
\hspace{-3cm}
+ P_L 
\left({1\over \not\! k{\bf 1}_8
-{\bf M}_C P_L-{\bf M}_C^\dag P_R}\right)_{\tilde H_d^\gamma \tilde B}P_L
\left({\bf h}_e\right)_{il}
\left(i\sigma_2\right)^{\gamma\beta}
\left({1\over k^2{\bf 1}_{12}-\widetilde{\bf M}_l^2}\right)_{\tilde L_l^\beta
\tilde L_j^{\dag\alpha}}\left(\frac{-g_1}{\sqrt{2}}\right)\\
&&
\hspace{-3cm}
+\sum_k P_L 
\left({1\over \not\! k{\bf 1}_8
-{\bf M}_C P_L-{\bf M}_C^\dag P_R}\right)_{\tilde H_d^\delta \tilde W^k}P_L
\left({\bf h}_e\right)_{il}
\left(i\sigma_2\right)^{\delta\gamma}
\left({1\over k^2{\bf 1}_{12}-\widetilde{\bf M}_l^2}\right)_{\tilde L_l^\gamma
\tilde L_j^{\dag\beta}}\ \left(\frac{g_2\sigma_k^{\beta\alpha}}{\sqrt{2}}\right)
\nonumber\\
&&
\hspace{-3cm}
\left. + P_L
\left(\frac{1}{\not\! k{\bf 1}_8-{\bf M}_C P_L-{\bf M}_C^\dag P_R}
\right)_{\tilde B \tilde H_d^\beta}
\left(\sqrt{2} g_1\right)
P_L\left({1\over k^2{\bf 1}_{12}-\widetilde{\bf M}_l^2}\right)_{\tilde E_i
\tilde E_l^\dag}
\left({\bf h}_e\right)_{lj}
\left(i\sigma\right)^{\beta\alpha}
\right] \ ,\nonumber
\end{eqnarray}
where  the  $9\times 9$  slepton  mass-squared matrix  $\widetilde{\bf
M}_l^2$ is displayed in Appendix~\ref{app_masses}.

Working  in  the  SHI  approximation,  we may  define  the  quantities
$\mbox{\boldmath$\Delta$}_e^{\Phi_i}     \equiv     {\bf     h}_e^{-1}
\mbox{\boldmath$\Delta$}{\bf       h}_e^{\Phi_i}$      in      analogy
with~(\ref{down_expansion}).    These  contributions  to   the  lepton
self-energy are given by
\begin{eqnarray}
\left<\mbox{\boldmath$\Delta$}_e^{\Phi_2}\right>_0
& = & 
{\bf 1}\ \frac{\alpha_1}{4\pi}\ \mu^\ast M_1^\ast
\ {\it I}\left(|M_1|^2, M_L^2, M_E^2\right)
+
{\bf 1}\ \frac{\alpha_1}{8\pi}\ \mu^\ast M_1^\ast
\ {\it I}\left(|M_1|^2, M_L^2, |\mu|^2\right)
\nonumber\\
& &
\hspace{-0.3cm}
-{\bf 1}\ \frac{3\alpha_2}{8\pi}\ \mu^\ast M_2^\ast
\ {\it I}\left(|M_2|^2, M_L^2, |\mu|^2\right)
-{\bf 1}\ \frac{\alpha_1}{4\pi}\ \mu^\ast M_1^\ast
\ {\it I}\left(|M_1|^2, M_E^2, |\mu|^2\right)\ ,\qquad\\
& &\nonumber\\
\left<\mbox{\boldmath$\Delta$}_e^{\Phi_1}\right>_0
& = & 
-{\bf 1}\ \frac{\alpha_1}{4\pi}\ A_e M_1^\ast
\ {\it I}\left(|M_1|^2, M_L^2, M_E^2\right)
-
{\bf 1}\ \frac{\alpha_1}{8\pi}
\ {\it B}_0\left(0, |M_1|^2, M_L^2\right)\nonumber\\
& &
+{\bf 1}\ \frac{3\alpha_2}{8\pi}
\ {\it B}_0\left(0, |M_2|^2, M_L^2\right)
+{\bf 1}\ \frac{\alpha_1}{4\pi}
\ {\it B}_0\left(0, |M_1|^2, M_E^2\right)\ ,\qquad
\end{eqnarray}
and  RG running  of the  soft SUSY-breaking  parameters leads  to a
shift which may be expressed as
\begin{eqnarray}
\label{ThresholdShift_e_phi2}
\Big<\mbox{\boldmath$\delta$}{\bf\Delta}^{\Phi_2}_e\Big>_0
& = &
\ \frac{\alpha_1}{4\pi}\ \mu^\ast M_1^\ast\ \left[
\mbox{\boldmath$\delta$}\widetilde{\bf M}^2_L
\ {\it K}\left(\widetilde{\bf M}^2_L, \widetilde{\bf M}^2_E, |M_1|^2\right)
+{\bf h}_e^{-1}\mbox{\boldmath$\delta$}\widetilde{\bf M}^2_E {\bf h}_e
\ {\it K}\left(\widetilde{\bf M}^2_E, \widetilde{\bf M}^2_L, |M_1|^2\right)
\right]
\nonumber\\
& &
+\frac{\alpha_1}{8\pi}\ \mu^\ast M_1^\ast\ \left[
\mbox{\boldmath$\delta$}\widetilde{\bf M}^2_L
\ {\it K}\left(\widetilde{\bf M}^2_L, |M_1|^2, |\mu|^2\right)
-2{\bf h}_e^\dag \mbox{\boldmath$\delta$}\widetilde{\bf M}^2_E {\bf h}_e
\ {\it K}\left(\widetilde{\bf M}^2_E, |M_1|^2, |\mu|^2\right)
\right]
\nonumber\\
& &
-\frac{3\alpha_2}{8\pi}\ \mu^\ast M_2^\ast\ 
\mbox{\boldmath$\delta$}\widetilde{\bf M}^2_L
\ {\it K}\left(\widetilde{\bf M}^2_L, |M_2|^2, |\mu|^2\right)\ ,
\\
& &\nonumber\\
\Big<\mbox{\boldmath$\delta$}{\bf\Delta}^{\Phi_1}_e\Big>_0
& = &
-\frac{\alpha_1}{4\pi}\ A_e M_1^\ast \left[
\mbox{\boldmath$\delta$}\widetilde{\bf M}^2_L
\ {\it K}\left(\widetilde{\bf M}^2_L, \widetilde{\bf M}^2_E, |M_1|^2\right)
+{\bf h}_e^{-1}\mbox{\boldmath$\delta$}\widetilde{\bf M}^2_E {\bf h}_e
\ {\it K}\left(\widetilde{\bf M}^2_E, \widetilde{\bf M}^2_L, |M_1|^2\right)
\right]
\nonumber\\
& &
-\frac{\alpha_1}{4\pi}\ \mbox{\boldmath$\delta$}{\bf a}_e M_1^\ast
\ {\it I}\left(\widetilde{\bf M}^2_L, \widetilde{\bf M}^2_E, |M_1|^2\right)
+\frac{3\alpha_2}{8\pi}\ \mbox{\boldmath$\delta$}\widetilde{\bf M}^2_L
\ {\it I}\left(\widetilde{\bf M}^2_L, |M_2|^2\right)
\nonumber\\
& &
+\frac{\alpha_1}{8\pi}\ \left[
\mbox{\boldmath$\delta$}\widetilde{\bf M}^2_L
\ {\it I}\left(\widetilde{\bf M}^2_L, |M_1|^2\right)
+2{\bf h}_e^{-1}\mbox{\boldmath$\delta$}\widetilde{\bf M}^2_E {\bf h}_e
\ {\it I}\left(\widetilde{\bf M}^2_E, |M_1|^2\right)
\right]\ .
\label{ThresholdShift_e_phi1}
\end{eqnarray}
where  the  flavour-non-universal   contributions  were  specified  in
(\ref{MFVshift}).

\subsection{Higgs Couplings in the Fermion Mass Basis}

The weak quark chiral states, $u^0_{L,R}$ and $d^0_{L,R}$, are related
to their respective mass eigenstates, $u_{L,R}$ and $d_{L,R}$, through
the unitary transformations:
\begin{eqnarray}
  \label{Utr}
&&
u^0_L\ =\ {\bf U}^Q_L\, u_L\,,\quad
d^0_L\ =\ {\bf U}^Q_L\, {\bf V}_d\, d_L\,,\quad
u^0_R\ =\ {\bf U}^u_R\, u_R\,,\quad
d^0_R\ =\ {\bf U}^d_R\, d_R\; \nonumber \\
&&
\nu^0_L\ =\ {\bf U}^L_L\, \nu_L\,,\quad \
e^0_L\ =\ {\bf U}^L_L\, {\bf V}_e\, e_L\,,\quad \,
e^0_R\ =\ {\bf U}^e_R\, e_R\,,
\end{eqnarray}
where ${\bf  U}^{Q,L}_L$, ${\bf U}^{u,d,e}_R$ are  $3\times 3$ unitary
matrices and ${\bf  V}_d$ and ${\bf V}_e^\dagger$ are  the CKM and the
PMNS mixing matrices, respectively.   The individual components of the
Higgs doublets $\Phi_{1,2}$ are given by
\begin{equation}
\Phi_{1,2}\ =\ \left(\! \begin{array}{c}
\phi^+_{1,2}\\
\frac{1}{\sqrt{2}}\, \Big(\, v_{1,2}\: +\: \phi_{1,2}\: +\: i a_{1,2}\,\Big)
\end{array}\! \right)\; .
\end{equation}
In the  CP-violating MSSM,  the weak-state Higgs  fields $\phi_{1,2}$,
$a_{1,2}$ and $\phi^-_{1,2}$ are  related to the neutral CP-mixed mass
eigenstates  $H_{1,2,3}$,  the  charged  Higgs  boson  $H^-$  and  the
would-be Goldstone bosons $G^0$ and $G^-$, associated with the $Z$ and
$W^-$ bosons, through~\cite{radCP,CPsuperH}:
\begin{eqnarray}
  \label{Hmix}
\phi_1 \!&=&\! O_{1i}\, H_i\,,\qquad\qquad\qquad\qquad\
\phi_2 \ =\ O_{2i}\, H_i\, ,\nonumber\\
a_1 \!&=&\! c_\beta\, G^0\: -\: s_\beta\, O_{3i}\, H_i\,,\qquad\quad\
a_2 \ =\ s_\beta\, G^0\: +\: c_\beta\, O_{3i}\, H_i\,,\nonumber\\
\phi^-_1 \!&=&\! c_\beta\, G^-\: -\: s_\beta\, H^-\,,\qquad\qquad
\phi^-_2 \ =\ s_\beta\, G^-\: +\: c_\beta\, H^-\; ,
\end{eqnarray}
where $s_\beta  \equiv \sin\beta$, $c_\beta \equiv  \cos\beta$ and $O$
is an orthogonal $3\times 3$ Higgs-boson-mixing matrix. 

The functional ${\bf h}_f^{-1} \Delta {\bf h}_f$ with $f=d,e,u$ can be
written as
\begin{eqnarray}
{\bf h}_{d,e}^{-1} \Delta {\bf h}_{d,e} &=&
(0,{\bf \Delta}_{d,e})\ +\ \sum_{i=1,2}\left(
{\bf \Delta}_{d,e}^{\phi_i^{-}}\,\phi_i^{-}\ , \
\frac{{\bf \Delta}_{d,e}^{\phi_i}}{\sqrt{2}}\,\phi_i\,+
\frac{{\bf \Delta}_{d,e}^{a_i}}{\sqrt{2}\,i}\,a_i
\right)\,, \nonumber \\
{\bf h}_u^{-1} \Delta {\bf h}_u &=&
(0,{\bf \Delta}_u)\ +\ \sum_{i=1,2}\left(
{\bf \Delta}_u^{\phi_i^{+}}\,\phi_i^{+}\ , \
\frac{{\bf \Delta}_u^{\phi_i}}{\sqrt{2}}\,\phi_i\,+
\frac{{\bf \Delta}_u^{a_i}}{\sqrt{2}\,i}\,a_i
\right)\,. 
\end{eqnarray}
The  $3\times  3$ matrices  ${\bf \Delta}_f$,
${\bf  \Delta}^{\phi^\pm_{i}}_f$,
${\bf  \Delta}^{\phi_{i}}_f$, and
${\bf  \Delta}^{a_{i}}_f$   are given by
\begin{eqnarray}
  \label{Deltad}
{\bf \Delta}_f &=&
\Big<\,{\bf h}_f^{-1} \Delta {\bf h}_f\,\Big>\,, 
\ \ \ \ \ \ \ \ \ \ \ \ \ \ \ \ \,
{\bf \Delta}^{\phi_{i}}_f \ =\ \sqrt{2}\;
\Big<\, \frac{\delta}{\delta \phi_{i}}\,
{\bf h}_f^{-1} \Delta {\bf h}_f\, \Big>\; ,\quad
\nonumber \\
{\bf \Delta}^{a_{i}}_f \ &=&\ i\,\sqrt{2}\;
\Big<\, \frac{\delta}{\delta a_{i}}\,
{\bf h}_f^{-1} \Delta {\bf h}_f\, \Big>\; ,\quad
{\bf \Delta}^{\phi^\pm_{i}}_f \ =\
\Big<\, \frac{\delta}{\delta \phi^\pm_{i}}\,
{\bf h}_f^{-1} \Delta {\bf h}_f\, \Big>\; ,
\end{eqnarray}
where $\langle\ldots\rangle$ indicates taking  the VEV of the enclosed
expression.   With these,  we may conveniently express  the general
flavour-changing (FC) effective Lagrangian for the interactions of the
neutral  and charged  Higgs fields  to  the up-  and down-type  quarks
$u,\ d$ and to the charged leptons in the following form:
\begin{eqnarray}
  \label{LeffFCNC}
{\cal L}_{\rm  FC} &=& -\,  \frac{g}{2 M_W}\ \Bigg[\,   \ \
H_i\; \bar{d}\,
\Big(\, \widehat{\bf  M}_d\, {\bf g}^L_{H_i\bar{d}d}\,  P_L\: +\: {\bf
g}^R_{H_i\bar{d}d}\,  \widehat{\bf M}_d\,  P_R\, \Big)\,  d\  +\ G^0\;
\bar{d}\,\Big(\,  \widehat{\bf M}_d\, i\gamma_5\,\Big)\,  d \nonumber\\
&& \ \ \ \ \ \ \ \ \ \ \ \ \
+\, H_i\; \bar{u}\, \Big(\, \widehat{\bf  M}_u\,
{\bf g}^L_{H_i\bar{u}u}\, P_L\: +\: {\bf g}^R_{H_i\bar{u}u}\,
\widehat{\bf  M}_u\, P_R\, \Big)\, u\ -\
G^0\,\bar{u}\,\Big(\, \widehat{\bf  M}_u\, i\gamma_5\,\Big)\, u\; \nonumber\\
&& \ \ \ \ \ \ \ \ \ \ \ \ \
+\, H_i\; \bar{e}\, \Big(\, \widehat{\bf  M}_e\,
{\bf g}^L_{H_i\bar{e}e}\, P_L\: +\: {\bf g}^R_{H_i\bar{e}e}\,
\widehat{\bf  M}_e\, P_R\, \Big)\, e\ +\
G^0\,\bar{e}\, \Big(\,\widehat{\bf  M}_e\, i\gamma_5\,\Big)\, e\;
\Bigg]\\ 
&&-\,  \frac{g}{\sqrt{2}   M_W}\  \Bigg[\,  
H^-\;   \bar{d}\,  \Big(\,
\widehat{\bf   M}_d\,  {\bf   g}^L_{H^-\bar{d}u}\,   P_L\:  +\:   {\bf
g}^R_{H^-\bar{d}u}\,  \widehat{\bf M}_u\,  P_R\,  \Big)\, u \,+\,
H^-\;   \bar{e}\,   \Big(\,
\widehat{\bf   M}_e\,  {\bf   g}^L_{H^-\bar{e}\nu}\,   P_L\, \Big)\,  \nu
\nonumber\\
&& \ \ \ \ \ \ \ \ \ \ \ \ \
+\ G^-\; \bar{d}\,  \Big(\, \widehat{\bf M}_d\,{\bf V}_d^\dagger P_L\:
-\:  {\bf V}_d^\dagger\,\widehat{\bf  M}_u\,  P_R\, \Big)\,  u\ \,+\,
G^-\; \bar{e}\,  \Big(\, \widehat{\bf M}_e\,{\bf V}_e^\dagger P_L\, \Big)\,  \nu\ 
+\  {\rm H.c.}\;\Bigg]\;, \nonumber
\end{eqnarray}
where  $\widehat{\bf  M}_{u,d,e}$   are  the  physical  diagonal  mass
matrices for the up- and down-type quarks and charged leptons, which
are related to the Yukawa coupling matrices by
\begin{equation}
  \label{hcond}
({\bf U}^{d,e}_R)^\dagger\, {\bf h}_{d,e}\, {\bf U}^{Q,L}_L\ =\
\frac{\sqrt{2}}{v_1}\, \widehat{\bf M}_{d,e}\, {\bf V}_{d,e}^\dagger\,
                                            {\bf R}^{-1}_{d,e}\,,\qquad
({\bf U}^{u}_R)^\dagger\, {\bf h}_u\, {\bf U}^Q_L\ =\
\frac{\sqrt{2}}{v_2}\, \widehat{\bf M}_u\, {\bf R}^{-1}_u\; ,
\end{equation}
with
\begin{eqnarray}
  \label{Rud}
{\bf R}_{d,e} &=& {\bf 1}\ +\ \frac{\sqrt{2}}{v_1}\, 
({\bf U}^{Q,L}_L)^\dagger\, {\bf \Delta}_{d,e}\,
{\bf U}^{Q,L}_L\,,\nonumber\\
{\bf R}_u &=& {\bf 1}\ +\ \frac{\sqrt{2}}{v_2}\, 
({\bf U}^{Q}_L)^\dagger\, {\bf \Delta}_{u}\,
{\bf U}^{Q}_L\ .
\end{eqnarray}
The neutral Higgs couplings to the down-type quarks and charged leptons
are given by
\begin{eqnarray}
  \label{gLd}
{\bf g}^L_{H_i\bar{f}f} &=& \frac{O_{1i}}{c_\beta}\;
{\bf V}_f^\dagger\, {\bf R}^{-1}_f\, \Big[ {\bf 1} + 
({\bf U}^F_L)^\dagger\,{\bf \Delta}^{\phi_1}_f\,{\bf U}^F_L \Big]\, {\bf V}_f\ +\
\frac{O_{2i}}{c_\beta}\; {\bf V}_f^\dagger\, {\bf R}^{-1}_f\,
({\bf U}^F_L)^\dagger\,{\bf \Delta}^{\phi_2}_f\, {\bf U}^F_L {\bf V}_f\nonumber\\
&&+\, iO_{3i}\, t_\beta\, {\bf V}_f^\dagger\, {\bf R}^{-1}_f\,
\Big[ {\bf 1} + ({\bf U}^F_L)^\dagger\,{\bf \Delta}^{a_1}_f\,{\bf
    U}^F_L - \frac{1}{t_\beta}\, 
({\bf U}^F_L)^\dagger\,{\bf \Delta}^{a_2}_f\,{\bf U}^F_L \Big]\, {\bf
  V}_f\; ,\\[3mm] 
  \label{gRd}
{\bf g}^R_{H_i\bar{f}f} &=& ({\bf g}^L_{H_i\bar{f}f})^\dagger\; ,
\end{eqnarray}
with $f=d,e$,  $F=Q,L$ and $t_\beta  \equiv \tan \beta$.   The neutral
Higgs couplings to the up-type quarks are given by
\begin{eqnarray}
  \label{gLu}
{\bf g}^L_{H_i\bar{u}u} &=& \frac{O_{1i}}{s_\beta}\;
{\bf R}^{-1}_u\,({\bf U}^Q_L)^\dagger\,{\bf \Delta}^{\phi_1}_u\,{\bf U}^Q_L\ +\
\frac{O_{2i}}{s_\beta}\; {\bf R}^{-1}_u\,
\Big[ {\bf 1} + ({\bf U}^Q_L)^\dagger\,{\bf \Delta}^{\phi_2}_u\,{\bf
    U}^Q_L \Big]\nonumber\\ 
&&+\, iO_{3i}\, t^{-1}_\beta\, {\bf R}^{-1}_u\,
\Big[ {\bf 1} - ({\bf U}^Q_L)^\dagger\,{\bf \Delta}^{a_2}_u\,{\bf U}^Q_L + t_\beta\,
({\bf U}^Q_L)^\dagger\,{\bf \Delta}^{a_1}_u\,{\bf U}^Q_L \Big]\; ,\\[3mm]
  \label{gRu}
{\bf g}^R_{H_i\bar{u}u} &=& ({\bf g}^L_{H_i\bar{u} u})^\dagger\; .
\end{eqnarray}
Finally, the coupling to the charged Higgs boson are given by
\begin{eqnarray}
  \label{gLud}
{\bf g}^L_{H^-\bar{d} u} &=&
-\, t_\beta\, {\bf V}_d^\dagger\, {\bf R}^{-1}_d\, \Big[
  {\bf 1} + ({\bf U}^Q_L)^\dagger\,{\bf \Delta}^{\phi^-_1}_d\,
{\bf  U}^Q_L \Big]\  +\
{\bf V}_d^\dagger\, {\bf R}^{-1}_d\,
({\bf U}^Q_L)^\dagger\,{\bf \Delta}^{\phi^-_2}_d\,{\bf U}^Q_L \; ,\\[3mm]
  \label{gRud}
{\bf g}^R_{H^-\bar{d} u} &=& -\, t^{-1}_\beta\, {\bf V}_d^\dagger\,
 \Big[  {\bf 1} + ({\bf U}^Q_L)^\dagger\,({\bf
  \Delta}^{\phi^+_2}_u )^\dagger\,{\bf U}^Q_L \Big]\, ({\bf R}^{-1}_u)^\dagger
\nonumber\\ 
&& +\
{\bf V}_d^\dagger\,({\bf U}^Q_L)^\dagger\,({\bf \Delta}^{\phi^+_1}_u )^\dagger\,
{\bf U}^Q_L\, ({\bf R}^{-1}_u )^\dagger\; ,\\[3mm]
  \label{gLenu}
{\bf g}^L_{H^-\bar{e} \nu} &=&
-\, t_\beta\, {\bf V}_e^\dagger\, {\bf R}^{-1}_e\, \Big[
  {\bf 1} + ({\bf U}^L_L)^\dagger\,{\bf \Delta}^{\phi^-_1}_e\,{\bf U}^L_L \Big]\  +\
{\bf V}_e^\dagger\, {\bf R}^{-1}_e\,
({\bf U}^L_L)^\dagger\,{\bf \Delta}^{\phi^-_2}_e\,{\bf U}^L_L\ .
\end{eqnarray}
Note that, in order to fully specify the Higgs couplings, one needs to
know    ${\bf    \Delta}_f$,    ${\bf    \Delta}_f^{\phi_i}$,    ${\bf
  \Delta}_f^{a_i}$, and ${\bf \Delta}_f^{\phi^\pm_i}$ with $i=1,2$ and
$f=d,u,e$,  as  well  as   the  rotation  matrices  ${\bf  U}^{Q,L}_L$
and~${\bf  V}_{d,e}$.   Hence,  the  analytic results  presented  here
generalise   those   given    in~\cite{MCPMFV,DP}   beyond   the   SHI
approximation in an arbitrary flavour basis.

\medskip

One may now exploit the properties of gauge- and flavour-covariance of
the  effective  functional ${\bf  h}_f^{-1}  \Delta {\bf  h}_f[\Phi_1,
  \Phi_2]$ with $f=e,d,u$ to obtain useful relations. Explicitly, they
should have the forms:
\begin{eqnarray}
\label{FandG}
{\bf h}_{e,d}^{-1} {\bf \Delta} {\bf h}_{e,d}[\Phi_1,  \Phi_2]\
&=&\ \Phi^\dagger_1\, {\bf F}_{e,d}\: +\: \Phi^\dagger_2\, {\bf G}_{e,d}\;,
\nonumber \\
{\bf h}_u^{-1} {\bf \Delta} {\bf h}_u[\Phi_1,  \Phi_2]\
&=&\ \Phi^T_1\, {\bf F}_u\: +\: \Phi^T_2\, {\bf G}_u\;,
\end{eqnarray}
where      ${\bf      F}_{e,d,u}      \Big(      \Phi^\dagger_1\Phi_1,
\Phi^\dagger_2\Phi_2,              \Phi^\dagger_1              \Phi_2,
\Phi^\dagger_2\Phi_1\Big)$     and      ${\bf     G}_{e,d,u}     \Big(
\Phi^\dagger_1\Phi_1,   \Phi^\dagger_2\Phi_2,  \Phi^\dagger_1  \Phi_2,
\Phi^\dagger_2\Phi_1\Big)$  are   calculable  $3\times  3$-dimensional
functionals  which in  general transform  as  ${\bf h}^\dagger_{e,d,u}
{\bf h}_{e,d,u}$ under the flavour rotations~(\ref{SUPERrot}).
In  the SHI approximation,  {\bf F}  and {\bf  G} are  simply constant
matrices, leading to the relations
\begin{eqnarray} 
{\bf \Delta}_{e,d}=\frac{v_1}{\sqrt{2}}\,{\bf F}^0_{e,d}
+\frac{v_2}{\sqrt{2}}\, {\bf G}^0_{e,d}\,,  &\ &
{\bf \Delta}_{u}=\frac{v_1}{\sqrt{2}}\,{\bf F}^0_u
+\frac{v_2}{\sqrt{2}}\, {\bf G}^0_{u} \,, 
\nonumber \\[3mm]
{\bf \Delta}_{e,d}^{\phi_1}={\bf \Delta}_{e,d}^{a_1}=
{\bf \Delta}_{e,d}^{\phi^-_1}= {\bf F}^0_{e,d} \,, &\ &
{\bf \Delta}_{u}^{\phi_1}=-{\bf \Delta}_{u}^{a_1}=
{\bf \Delta}_{u}^{\phi^+_1}= {\bf F}^0_{u} \,, \\[3mm]
{\bf \Delta}_{e,d}^{\phi_2}={\bf \Delta}_{e,d}^{a_2}=
{\bf \Delta}_{e,d}^{\phi^-_2}= {\bf G}^0_{e,d} \,, &\ &
{\bf \Delta}_{u}^{\phi_2}=-{\bf \Delta}_{u}^{a_2}=
{\bf \Delta}_{u}^{\phi^+_2}= {\bf G}^0_{u} \, .\nonumber
\end{eqnarray}
Including the RG-induced  flavour-violating terms,  the constant
matrices {\bf F} and {\bf G} take on the form:
\begin{equation} 
{\bf F}^0_{e,d,u} = 
\Big< {\bf \Delta}_{e,d,u}^{\Phi_1}+\mbox{\boldmath$\delta\Delta$}_{e,d,u}^{\Phi_1}
\Big>_0\; , \qquad
{\bf G}^0_{e,d,u} = 
\Big< {\bf \Delta}_{e,d,u}^{\Phi_2}+\mbox{\boldmath$\delta\Delta$}_{e,d,u}^{\Phi_2}
\Big>_0\; ,
\end{equation}
where  the  flavour-non-universal  terms  are  given  by  the  set  of
equations                             (\ref{ThresholdShift_down_phi2}),
(\ref{ThresholdShift_down_phi1}),       (\ref{ThresholdShift_up_phi2}),
(\ref{ThresholdShift_up_phi1}),    (\ref{ThresholdShift_e_phi2})   and
(\ref{ThresholdShift_e_phi1}).  In  the next section,  we will present
numerical estimates in the SHI approximation, including the RG-induced
flavour-non-universal contributions to the threshold corrections.

\subsection{RG Effects on FCNC Higgs-Boson Couplings}

We may now illustrate the significance of the results presented in the
previous section, by analyzing the variations of the FCNC couplings of
the  neutral  Higgs  bosons   to  down-type  quarks  as  functions  of
$\tan\beta$ in the  same MCPMFV scenario (\ref{eq:cpsps1a}) introduced
previously~\footnote{We  recall  that, in  this  scenario, the  heavier
  neutral Higgs bosons $H_{2,3}$  are almost degenerate, whereas $H_1$
  is   significantly    lighter.}.    

\begin{figure}[thp]
\hspace{ 0.0cm}
\centerline{\epsfig{figure=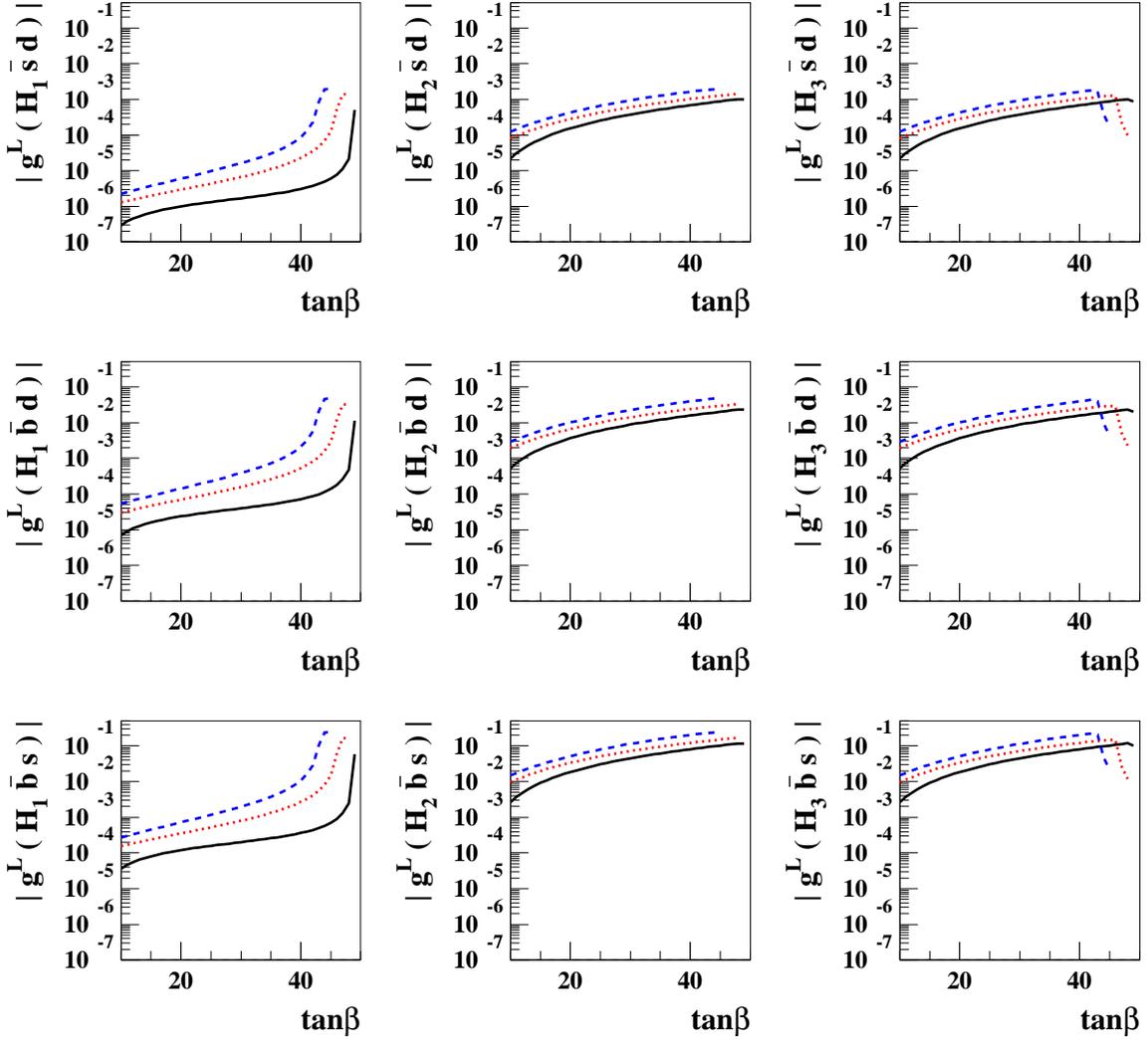,height=16cm,width=16cm}}
\vspace{-0.9cm}
\caption{\it  In the upper, middle, and lower panels
we display the variation with $\tan \beta$ of
the flavour-non-diagonal couplings 
$|g^L_{H_{i}{\bar s}d}|$,
$|g^L_{H_{i}{\bar b}d}|$, and
$|g^L_{H_{i}{\bar b}s}|$, respectively,
of the neutral Higgs bosons 
$H_1$ (left column), $H_2$ (middle column), and $H_3$ (right column).
In each panel, the lines are the same as in Fig.~\ref{fig:cmq2}.
The input MCPMFV SUSY-breaking parameters are taken 
to be the same as in (\protect\ref{eq:cpsps1a}).
}
\label{fig:fcnc.coupl}
\end{figure}

To start with, we  display in Fig.~\ref{fig:fcnc.coupl} the variations
with  $\tan\beta$ of  the couplings  $|g^L(H_{1,2,3}  \bar{s}d)|$ (top
row),  $|g^L(H_{1,2,3} \bar{b}d)|$  (middle  row) and  $|g^L(H_{1,2,3}
\bar{b}s)|$,   for   various   values   of  $\Phi_M$   as   given   in
Fig.~\ref{fig:cmq2}.  We see that, for each of the Higgs particles, we
have  the   expected  hierarchy  across   fermion  generations,  i.e.,
$|g^L(H_{1,2,3}   \bar{s}d)|/|g^L(H_{1,2,3}  \bar{b}d)|   \sim  |({\bf
V}_d)_{ts}|  \sim 0.04$ and  $|g^L(H_{1,2,3} \bar{b}d)|/|g^L(H_{1,2,3}
\bar{b}s)|\sim |({\bf V}_d)_{td}/({\bf  V}_d)_{ts}| \sim 0.2$.  Again,
we  observe  that  for  $\Phi_M=180^\circ$, gluino  threshold  effects
increase  the value  of the  threshold matrix  ${\bf  R}_d^{-1}$, thus
giving rise to larger FCNC Higgs-boson couplings.

\begin{figure}[thp]
\hspace{ 0.0cm}
\centerline{\epsfig{figure=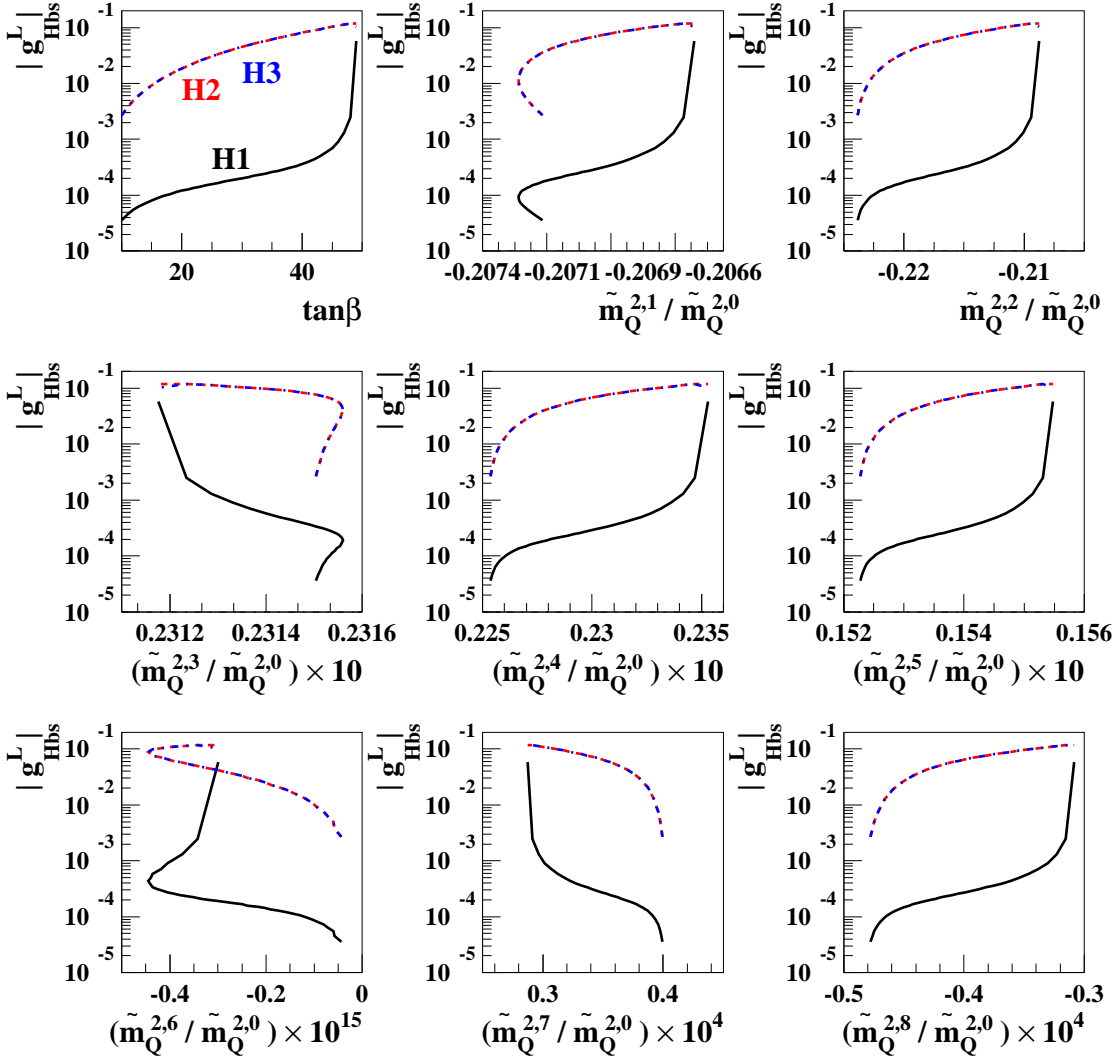,height=16cm,width=16cm}}
\vspace{-0.9cm}
\caption{\it In the top left panel we display the variation with $\tan
  \beta$ of the flavour-non-diagonal couplings $|g^L_{H_{i}{\bar
      b}s}|$ of the neutral Higgs bosons $H_{1, 2, 3}$ taking
  $\Phi_M=0^\circ$.  In the other panels we display the correlations
  of these couplings with the ratios
  $\widetilde{m}^{2,I}_Q/\widetilde{m}^{2,0}_Q$ for the same value of
  $\Phi_M=0^\circ$.  The input MCPMFV SUSY-breaking parameters are
  taken to be the same as in (\protect\ref{eq:cpsps1a}).  }
\label{fig:gbs.cmq2}
\end{figure}

In  the  top  left  panel  of  Fig.~\ref{fig:gbs.cmq2},  we  show  the
dependence of the couplings $|g^L(H_{1,2,3} \bar{b}s)|$ on $\tan\beta$
for  $\Phi_M=0$, whilst  the  remaining panels  show the  correlations
between  this   coupling  and  each  of   the  expansion  coefficients
$\widetilde{m}_Q^{2,I}/\widetilde{m}_Q^{2,0}$  for the same  values of
$\tan\beta$.    As   expected    from   the   results   displayed   in
Fig.~\ref{fig:cmq2},  the variation in  the coefficients  is extremely
small.

\begin{figure}[thp]
\hspace{ 0.0cm}
\centerline{\epsfig{figure=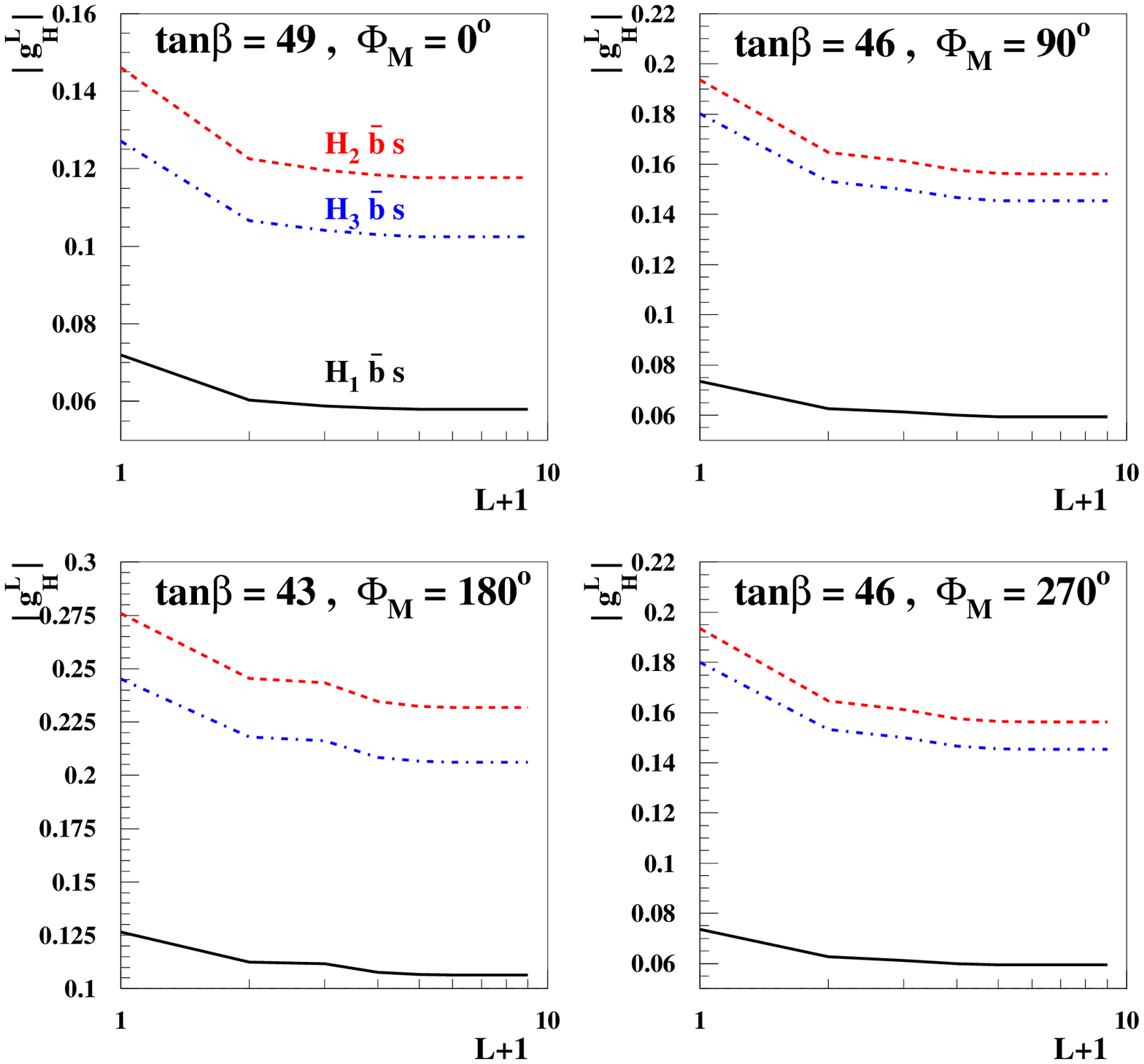,height=16cm,width=16cm}}
\vspace{-0.9cm}
\caption{\it In the top left panel we display the variation with the
  level $L$ of the flavour-non-diagonal couplings $|g^L_{H_{i}{\bar
      b}s}|$ of the neutral Higgs bosons $H_{1, 2, 3}$ taking
  $\tan\beta=49$ and $\Phi_M=0^\circ$.  The level $L$ is defined in
  the sum: $\widetilde{\bf M}^2_{Q,U,D} =
  \sum\limits_{I=0}^L\;\widetilde{m}^{2,I}_{Q,U,D}\; {\bf
    H}^{Q,U,D}_I$ and ${\bf a}_{u,d} = \sum\limits_{I=0}^L\,
  a^I_{u,d}\, {\bf h}_{u,d}\, {\bf H}^Q_I$.
For example, $L=0$ implies that the couplings $|g^L_{H_{i}{\bar b}s}|$
are calculated   
only including the zero-th term in the expansions of the
matrices $\widetilde{\bf M}^2_{Q,U,D}$ and ${\bf a}_{u,d}$
while $L=8$ means including all 9 terms in the expansions.
In the other panels we display the variations
for other values of $\tan\beta$ and $\Phi_M$ as shown.
The input MCPMFV SUSY-breaking parameters are taken
to be the same as in (\protect\ref{eq:cpsps1a}).
}
\label{fig:gbs.coupl}
\end{figure}

Finally, in  Fig.~\ref{fig:gbs.coupl} we illustrate the  impact of the
individual flavour-violating  components induced by RG  running on the
FCNC Higgs-boson couplings $|g^L(H_{1,2,3} \bar{b}s)|$.  Specifically,
we plot the couplings $|g^L(H_{1,2,3} \bar{b}s)|$ against the ``level"
$L$, defined by limiting the sums
\begin{eqnarray}
\widetilde{\bf M}^{2\, (L)}_{Q,U,D} & = & 
\sum\limits_{I=0}^L\;\widetilde{m}^{2,I}_{Q,U,D}\; {\bf H}^{Q,U,D}_I\ ,
\nonumber\\
{\bf a}^{(L)}_{u,d} & = & \sum\limits_{I=0}^L\, a^I_{u,d}\, {\bf
  h}_{u,d}\, {\bf H}^Q_I\ . 
\end{eqnarray}
The level $L=0$ indicates  that the couplings are calculated including
only  the first  coefficient in  the expansion  of the  soft matrices,
i.e.~the flavour-singlet component $\propto  {\bf 1}_3$.  On the other
hand, the highest level $L=8$ gives the complete reconstruction of the
soft   SUSY-breaking   matrices    in   a   flavour-basis   expansion,
i.e.~$\widetilde{\bf   M}^{2\,   (8)}_{Q,U,D}  \equiv   \widetilde{\bf
  M}^2_{Q,U,D}$  and  ${\bf   a}^{(8)}_{u,d}  \equiv  {\bf  a}_{u,d}$.

We  also display in  Fig.~\ref{fig:gbs.coupl} numerical  estimates for
various values of $\tan\beta$  and $\Phi_M$.  In~all different panels,
we   observe   that  the   effective   Higgs-boson  couplings   remain
flavour-violating,  even  if   the  soft  SUSY-breaking  matrices  are
flavour-singlet.   In  this   case,  the  Higgsino-mediated  threshold
correction and the 2HDM contribution,  both of which are $\propto {\bf
h}^\dagger_u  {\bf  h}_u$,  remain  the  dominant  source  of  flavour
violation when all the soft SUSY-breaking matrices are proportional to
${\bf 1}_3$.  Interestingly  enough, the RG-generated flavour-changing
terms in the soft matrices are found to produce a cancellation against
the leading level $L=0$ terms.  This screening phenomenon is typically
at the level of $\sim 20\%$ of the Higgsino contribution.  The largest
non-flavour-singlet   contribution   is  due   to   the  level   $L=1$
coefficients,   although  the  $L=2,3$   coefficients  also   lead  to
appreciable effects  for some values of $\Phi_M$.   Thus, our analysis
shows  that  the  non-singlet  flavour components  within  MCPMFV-type
scenarios can modify the strength of the FCNC Higgs-boson couplings in
a phenomenologically relevant way.

\section{Conclusions}

We have presented a new  geometric method for the flavour decomposition
of a general soft SUSY-breaking sector in the MSSM.  We have shown how
the up-  and down-type quark  Yukawa-coupling matrices may be  used to
furnish a complete  geometric basis for the flavour  space of the soft
SUSY-breaking squark  mass matrices  and trilinear couplings.   Such a
decomposition  in the leptonic  sector would require  the extension  of the
MSSM by right-handed neutrinos and the introduction of neutrino Yukawa
couplings.

The MFV decomposition  developed here is valid except  possibly in the
limit where the effective Yukawa couplings are rendered real after the
inclusion of threshold corrections. In this case, an alternative basis
must be chosen;  the tree-level quark Yukawa couplings  are a suitable
choice.  Conversely, if the tree-level  Yukawa couplings are real in a
MSSM scenario  with soft  CP violation, then  RG effects  alone cannot
alter this fact.   In such a scenario, the  effective Yukawa couplings
are expected to acquire a  non-trivial KM phase induced by the complex
soft SUSY-breaking  parameters and they  will thus provide  a complete
basis for flavour decomposition.

It is obvious that threshold  corrections to the couplings of the MSSM
Higgs bosons to fermions play a  central role to the construction of a
non-degenerate flavour  geometry as developed here.   In this context,
we have presented the complete one-loop threshold corrections, showing
explicitly how non-trivial flavour  geometry of the soft SUSY-breaking
matrices  can become  an additional  source  for FCNC  effects on  the
Higgs-boson   couplings  to   fermions.    We  have   then  used   the
flavour-space  decomposition   to  examine  the   behaviour  of  these
couplings.   We have  observed that  flavour violation  due to  the RG
evolution  of the  soft SUSY-breaking  matrices  from the  GUT to  the
electroweak scale leads to a screening phenomenon of order $\sim 20\%$
for   the   effective   Higgs-fermion   vertices,   which   could   be
phenomenologically relevant.

In this  paper we have applied  our flavour geometric  approach to one
process,  namely  to  the  Higgs-boson  FCNC  couplings  to  fermions.
Non-trivial flavour structures amongst the soft SUSY-breaking matrices
are  known to  contribute to  a number  of processes  at  the one-loop
level,   such   as   $b\to   s\gamma$   decays   and   neutral   meson
mixing~\cite{Buras}.  Applying  the framework developed  here to these
processes would lead  to constraints on the magnitudes  of the various
coefficients which would in turn have implications for the validity of
the Minimal Flavour Violation hypothesis.

\section*{Acknowledgements}
The  work of  AP was  supported in  part by  the STFC  research grant:
PP/D000157/1.   RNH~acknowledges support from the Japan Society for
the Promotion of Science.

\newpage
\begin{appendix}
\setcounter{equation}{0}
\def\theequation{\Alph{section}.\arabic{equation}}

\section{Background Higgs-Field-Dependent Mass Matrices}
\label{app_masses}

Here we  collect explicit tree-level  forms for all the  mass matrices
which        appear        in        the        calculations        of
Section~\ref{threshold_corrections}.

\subsection{Squark mass-squared matrix}

The $12\times 12$ squark mass-squared matrix $\widetilde{\bf M}^2$ may be
expressed as \cite{MCPMFV}
\begin{equation}
\label{squarkmassmatrix}
\widetilde{\bf M}^2 [\Phi_1,\Phi_2,S]=
\left(\begin{array} {ccc}
\left(\widetilde{\bf M}^2\right)_{\tilde Q^\dag \tilde Q} &
\left(\widetilde{\bf M}^2\right)_{\tilde Q^\dag \tilde U} &
\left(\widetilde{\bf M}^2\right)_{\tilde Q^\dag \tilde D} \\ 
\left(\widetilde{\bf M}^2\right)_{\tilde U^\dag \tilde Q} &
\left(\widetilde{\bf M}^2\right)_{\tilde U^\dag \tilde U} &
\left(\widetilde{\bf M}^2\right)_{\tilde U^\dag \tilde D} \\
\left(\widetilde{\bf M}^2\right)_{\tilde D^\dag \tilde Q} &
\left(\widetilde{\bf M}^2\right)_{\tilde D^\dag \tilde U} &
\left(\widetilde{\bf M}^2\right)_{\tilde D^\dag \tilde D}
\end{array}\right)_{ij}\ \ ,
\end{equation}
with
\begin{eqnarray}
 \label{squarkmatrixelements}
\left(\widetilde{\bf M}^2\right)_{\tilde Q_i^\dag \tilde Q_j}
& = &
\left(\widetilde{\bf M}_Q^2\right)_{ij}{\bf 1}_2
+\left({\bf h}^\dag_d{\bf h}_d\right)_{ij}\Phi_1\Phi_1^\dag
+\left({\bf h}^\dag_u{\bf h}_u\right)_{ij}
\left(\Phi_2^\dag\Phi_2{\bf 1}_2-\Phi_2\Phi_2^\dag\right)\nonumber\\
&&
-\frac{1}{2}\delta_{ij} g_2^2\left(\Phi_1\Phi_1^\dag-\Phi_2\Phi_2^\dag\right)
+\delta_{ij}\left(\frac{1}{4}g_2^2-\frac{1}{12}{g_1}^2\right)
\left(\Phi_1^\dag\Phi_1-\Phi_2^\dag\Phi_2\right){\bf 1}_2,
\nonumber\\
\left(\widetilde{\bf M}^2\right)_{\tilde U^\dag_i\tilde Q_j}
& = &
\left(\widetilde{\bf M}^2\right)^\dag_{\tilde Q^\dag_j\tilde U_i}  =
-\left({\bf a}_u\right)_{ij}\Phi_2^T i\sigma_2
+\left({\bf h}_u\right)_{ij}\mu^\ast \Phi_1^T i\sigma_2 ,
\nonumber\\
\left(\widetilde{\bf M}^2\right)_{\tilde D^\dag_i\tilde Q_j}
& = &
\left(\widetilde{\bf M}^2\right)^\dag_{\tilde Q^\dag_j\tilde D_i}  =
+\left({\bf a}_d\right)_{ij}\Phi_1^\dag
-\left({\bf h}_d\right)_{ij}\mu^\ast \Phi_2^\dag ,
\nonumber\\
\left(\widetilde{\bf M}^2\right)_{\tilde U_i^\dag \tilde U_j}
& = &
\left(\widetilde{\bf M}^2_U\right)_{ij}
+\left({\bf h}_u{\bf h}_u^\dag\right)_{ij}\Phi_2^\dag\Phi_2
+\frac{1}{3}\delta_{ij} {g_1}^2\left(\Phi_1^\dag\Phi_1-\Phi_2^\dag\Phi_2
\right) ,
\nonumber\\
\left(\widetilde{\bf M}^2\right)_{\tilde D_i^\dag \tilde D_j}
& = &
\left(\widetilde{\bf M}^2_D\right)_{ij}
+\left({\bf h}_d{\bf h}_d^\dag\right)_{ij}\Phi_1^\dag\Phi_1
-\frac{1}{6}\delta_{ij} {g_1}^2\left(\Phi_1^\dag\Phi_1-\Phi_2^\dag\Phi_2
\right) ,
\nonumber\\
\left(\widetilde{\bf M}^2\right)_{\tilde U_i^\dag \tilde D_j}
& = &
\left(\widetilde{\bf M}^2\right)^\dag_{\tilde D_j^\dag \tilde U_i} =
\left({\bf h}_u{\bf h}_d^\dag\right)_{ij}\Phi_1^T i\sigma\Phi_2\ 
\end{eqnarray}
where   ${\widetilde{\bf   M}_Q^2},\   {\widetilde{\bf  M}_U^2}$   and
${\widetilde{\bf  M}_D^2}$  are  the  soft SUSY-breaking  squark  mass
matrices and  $\mu$ is the coefficient  of the Higgs  bilinear term of
the MSSM superpotential.

\subsection{Chargino-Neutralino mass matrix}

The $8\times 8$ chargino-neutralino mass matrix ${\bf M}_C$ which appears
in~(\ref{complete_down}) is given by
\begin{equation}
\label{charginomassmatrix}
{\bf M}_C =
\left(\begin{array} {cccc}
M_1 & 0 & \frac{1}{\sqrt 2} g_1 \Phi_2^\dag &
\frac{1}{\sqrt 2} g_1\Phi_1^T (i\sigma_2)\\
0 & M_2{\bf 1}_3 & \frac{1}{\sqrt 2} g_2 \Phi_2^\dag \sigma_i
& -\frac{1}{\sqrt 2} g_2 \Phi_1^T (i\sigma_2) \sigma_i\\
\frac{1}{\sqrt 2} g_1 \Phi_2^\ast & \frac{1}{\sqrt 2} g_2
\sigma_i^T\Phi_2^\ast & {\bf 0}_2 & \mu (i\sigma_2)\\
-\frac{1}{\sqrt 2}(i\sigma_2) g_1 \Phi_1 & \frac{1}{\sqrt 2} g_2 \sigma_i^T
(i\sigma_2)\Phi_1 & -\mu (i\sigma_2) & {\bf 0}_2
\end{array} \right)\ .
\end{equation}

\subsection{Slepton mass-squared matrix}

The $9\times9$
slepton mass-squared matrix $\widetilde{\bf M}_l^2$ may be expressed as
\begin{equation}
\widetilde{\bf M}_l^2
=
\left(\begin{array}{cc}
\left(\widetilde{\bf M}^2\right)_{\tilde L^\dag \tilde L} &
\left(\widetilde{\bf M}^2\right)_{\tilde L^\dag \tilde E} \\
\left(\widetilde{\bf M}^2\right)_{\tilde E^\dag \tilde L} &
\left(\widetilde{\bf M}^2\right)_{\tilde E^\dag \tilde E} \end{array}
\right)\ ,
\end{equation}
with
\begin{eqnarray}
\left(\widetilde{\bf M}_l^2\right)_{\tilde L^\dag\tilde L} & = &
\left(\widetilde{\bf M}_L^2\right)_{ij}{\bf 1}_2
+\left({\bf h}^\dag_e{\bf h}_e\right)_{ij}\Phi_1\Phi_1^\dag
-\frac{1}{2}\delta_{ij} g_2^2\left(\Phi_1\Phi_1^\dag-\Phi_2\Phi_2^\dag\right)
\nonumber\\
&&
+\frac{1}{4}\delta_{ij}\left(g_2^2+{g_1}^2\right)
\left(\Phi_1^\dag\Phi_1-\Phi_2^\dag\Phi_2\right){\bf 1}_2\,
\nonumber\\
\left(\widetilde{\bf M}_l^2\right)_{\tilde E^\dag\tilde L} & = &
\left(\widetilde{\bf M}_l^2\right)_{\tilde L^\dag\tilde E}^\dag =
+\left({\bf a}_e\right)_{ij}\Phi_1^\dag
-\left({\bf h}_e\right)_{ij}\mu^\ast \Phi_2^\dag\ ,
\nonumber\\
\left(\widetilde{\bf M}_l^2\right)_{\tilde E^\dag\tilde E}& = &
\left(\widetilde{\bf M}^2_E\right)_{ij}
+\left({\bf h}_e{\bf h}_e^\dag\right)_{ij}\Phi_1^\dag\Phi_1
-\frac{1}{2}\delta_{ij} {g_1}^2\left(\Phi_1^\dag\Phi_1-\Phi_2^\dag\Phi_2
\right)\ .
\end{eqnarray}
Here ${\widetilde{\bf M}_L^2}$ and ${\widetilde{\bf M}_E^2}$ are the
soft SUSY-breaking slepton mass matrices.

\subsection{Higgs-boson mass-squared matrix}

The $8\times 8$ Higgs-boson mass-squared matrix ${\bf M}_H^2$ is given at tree
level in the basis $\left(\Phi_1,\Phi_2,\Phi^\ast_1,\Phi_2^\ast\right)^T$ by
\begin{equation}
{\bf M}_H^2\left[\Phi_1,\Phi_2\right]
=
\left(\begin{array}{cccc}
\left({\bf M}_H^2\right)_{\Phi_1^\dag\Phi_1} &
\left({\bf M}_H^2\right)_{\Phi_1^\dag\Phi_2} &
\left({\bf M}_H^2\right)_{\Phi_1^\dag\Phi_1^\ast} &
\left({\bf M}_H^2\right)_{\Phi_1^\dag\Phi_2^\ast} \\
\left({\bf M}_H^2\right)_{\Phi_2^\dag\Phi_1} &
\left({\bf M}_H^2\right)_{\Phi_2^\dag\Phi_2} &
\left({\bf M}_H^2\right)_{\Phi_2^\dag\Phi_1^\ast} &
\left({\bf M}_H^2\right)_{\Phi_2^\dag\Phi_2^\ast} \\
\left({\bf M}_H^2\right)_{\Phi_1^T\Phi_1} &
\left({\bf M}_H^2\right)_{\Phi_1^T\Phi_2} &
\left({\bf M}_H^2\right)_{\Phi_1^T\Phi_1^\ast} &
\left({\bf M}_H^2\right)_{\Phi_1^T\Phi_2^\ast} \\
\left({\bf M}_H^2\right)_{\Phi_2^T\Phi_1} &
\left({\bf M}_H^2\right)_{\Phi_2^T\Phi_2} &
\left({\bf M}_H^2\right)_{\Phi_2^T\Phi_1^\ast} &
\left({\bf M}_H^2\right)_{\Phi_2^T\Phi_2^\ast}
\end{array}\right)\ ,
\end{equation}
where
\begin{eqnarray}
\left({\bf M}_H^2\right)_{\Phi_1^\dag\Phi_1}
& = &
\left({\bf M}_H^2\right)_{\Phi_1^T\Phi_1^\ast}^T
\nonumber\\
& = &
\left(M^2_{H_d} + |\mu|^2 + \frac{g_1^2+g_2^2}{2}\Phi_1^\dag\Phi_1
+\frac{g_2^2-g_1^2}{4}\Phi_2^\dag\Phi_2\right){\bf 1}_2
-\frac{g_2^2}{2}\Phi_2\Phi_2^\dag\ ,
\nonumber\\
\left({\bf M}_H^2\right)_{\Phi_2^\dag\Phi_2}
& = &
\left({\bf M}_H^2\right)_{\Phi_2^T\Phi_2^\ast}^T
\nonumber\\
& = &
\left(M^2_{H_u} + |\mu|^2 + \frac{g_1^2+g_2^2}{2}\Phi_2^\dag\Phi_2
+\frac{g_2^2-g_1^2}{4}\Phi_1^\dag\Phi_1\right){\bf 1}_2
-\frac{g_2^2}{2}\Phi_1\Phi_1^\dag\ ,
\nonumber\\
\left({\bf M}_H^2\right)_{\Phi_1^T\Phi_2}
& = &
\left({\bf M}_H^2\right)_{\Phi_2^T\Phi_1}^T
=
\left({\bf M}_H^2\right)_{\Phi_2^\dag\Phi_1^\ast}^\dag
=
\left({\bf M}_H^2\right)_{\Phi_1^\dag\Phi_2^\ast}^\ast
\nonumber\\
& = &
\frac{g_2^2-g_1^2}{4}\Phi_1^\ast\Phi_2^\dag
-\frac{g_2^2}{2}\Phi_2^\ast\Phi_1^\dag\ ,
\nonumber\\
\left({\bf M}_H^2\right)_{\Phi_1^\dag\Phi_2} & = &
\left({\bf M}_H^2\right)^\dag_{\Phi_2^\dag\Phi_1}
=
\left({\bf M}_H^2\right)_{\Phi_2^T\Phi_1}^T
=
\left({\bf M}_H^2\right)_{\Phi_1^T\Phi_2^\ast}^\ast
\nonumber\\
& = &
-\left(B\mu + \frac{g_2^2}{2}\Phi_2^\dag\Phi_1\right){\bf 1}_2
+\frac{g_2^2-g_1^2}{4}\Phi_1\Phi_2^\dag\ ,
\nonumber\\
\left({\bf M}_H^2\right)_{\Phi_1^T\Phi_1}
& = &
\left({\bf M}_H^2\right)_{\Phi_1^\dag\Phi_1^\ast}^\ast
=
\frac{g_2^2+g_1^2}{2}\Phi_1^\ast\Phi_1^\dag\ ,
\nonumber\\
\left({\bf M}_H^2\right)_{\Phi_2^T\Phi_2}
& = &
\left({\bf M}_H^2\right)_{\Phi_2^\dag\Phi_2^\ast}^\ast
=
\frac{g_2^2+g_1^2}{2}\Phi_2^\ast\Phi_2^\dag\ .
\end{eqnarray}

\subsection{Quark mass matrix}

The $6\times 6$ quark mass matrix ${\bf M}_q$ is given by
\begin{equation}
{\bf M}_q\left[\Phi_1,\Phi_2\right]
=
\left(\begin{array}{c}
\left({\bf M}_q\right)_{u_iQ_j}\\
\left({\bf M}_q\right)_{d_iQ_j}
\end{array}\right)
=
\left(\begin{array}{c}
\left({\bf h}_u\right)_{ij} \Phi_2^T\left(-i\sigma_2\right)\\
\left({\bf h}_d\right)_{ij} \Phi_1^\dag
\end{array}\right)\ .
\end{equation}

\end{appendix}

\end{document}